\RequirePackage{fix-cm}
\documentclass[smallextended]{svjour3}       % onecolumn (second format)
\smartqed  % flush right qed marks, e.g. at end of proof
\usepackage{graphicx}

\usepackage[numbers]{natbib}
\usepackage{amsmath}
\usepackage{multirow}

\journalname{International Journal of Advanced Manufacturing Technology}

\begin{document}

\title{Accurate prediction of machining feedrate and cycle time considering interpolator dynamics}     

\titlerunning{Feedrate Prediction and Cycle Time Estimation} 

\author{Rob Ward         \and
        Burak Sencer \and
        Bryn Jones \and
        Erdem Ozturk
}

%\authorrunning{Short form of author list} % if too long for running head

\institute{R. Ward \at
              Industrial Doctorate Centre in Machining Science, Advanced Manufacturing Research Centre, University of Sheffield, Rotherham, S60 5TZ, UK \\
              \email{r.ward@amrc.co.uk}           %  \\
%             \emph{Present address:} of F. Author  %  if needed
           \and
           B. Sencer \at
              Mechanical Engineering Department, Oregon State University, Corvallis, OR, USA \\
              \email{burak.sencer@oregonstate.edu}
           \and
           B. Jones \at
           Department of Automatic Control and Systems Engineering, University of Sheffield, Sheffield, S1 3JD \\
           \email{b.l.jones@sheffield.ac.uk}
          \and
          E. Ozturk \at
          Advanced Manufacturing Research Centre, University of Sheffield, Rotherham, S60 5TZ, UK
          \email{e.ozturk@amrc.co.uk}
}

\date{Received: date / Accepted: date}
% The correct dates will be entered by the editor

\maketitle

\begin{abstract}
This paper presents an accurate machining feedrate prediction technique by modeling the trajectory generation behaviour of modern CNC machine tools. Typically, CAM systems simulate
machines’ motion based on the commanded feedrate and the path geometry. Such
approach does not consider the feed planning and interpolation strategy of the machine’s numerical control (NC) system. In this study, trajectory generation behaviour of the NC system is modelled and accurate cycle time prediction for complex machining toolpaths is realized. NC system’s linear interpolation dynamics and commanded axis kinematic profiles are predicted by using Finite Impulse Response (FIR) based low-pass filters. The corner blending behaviour during non-stop interpolation of linear segments is modeled, and for the first time, the minimum cornering feedrate, that satisfies both the tolerance and machining constraints, has been calculated analytically for 3-axis toolpaths of any geometry. The proposed method is applied to 4 different case studies including complex machining tool-paths. Experimental validations show actual cycle times can be estimated with $>$90\% accuracy, greatly outperforming CAM-based predictions. It is expected that the proposed approach will help improve the accuracy of virtual machining models and support businesses decision making when costing machining processes.


\end{abstract}
Incorrect cycle times may cost businesses during contract stages.

\maketitle

\section{Introduction}\label{Section:Introduction}
  
With the introduction of concepts like virtual manufacturing \cite{Altintas2016} and digital twins \cite{Armendia2019}, building process models and predicting actual machining process conditions in the computer environment has become paramount in attaining higher productivity and throughput in today's manufacturing. For example, accurate machining cycle time prediction is vital for industry during the quotation process to ensure achievable and profitable contracts. The prediction models and generation of accurate digital twins is a collective modeling effort which requires both detailed modelling of the process as well as the dynamic machine behaviour. Considering the machining processes, current literature provides accurate models to predict milling process physics \cite{LayeghK.2012,Berglind2017}. Nevertheless, when applied in practice, these models show large discrepancies from the actual process behaviour. 

One reason can be identified as the influence of the machine tool drive dynamics. In particular, the behaviour of the Numerical Control (NC) plays a key role. Trajectory generation (interpolation) algorithms embedded in the NC system, control the feedrate profile, which is a key input for machining process models. For example, contouring (positioning) errors alter tool engagements \cite{Sencer2009a} which lead to inaccurate force predictions \cite{Altintas2014b}. Thus, in order to accurately develop realistic digital twins for machining processes, the feedrate profile generated by the NC system of a machine tool must be accurately predicted. This paper deals with modeling and prediction of interpolator dynamics of modern NC systems to accurately estimate machining cycle times and cutting forces along complex parts. 

\begin{figure}
	\centering
	\includegraphics[width=\textwidth]{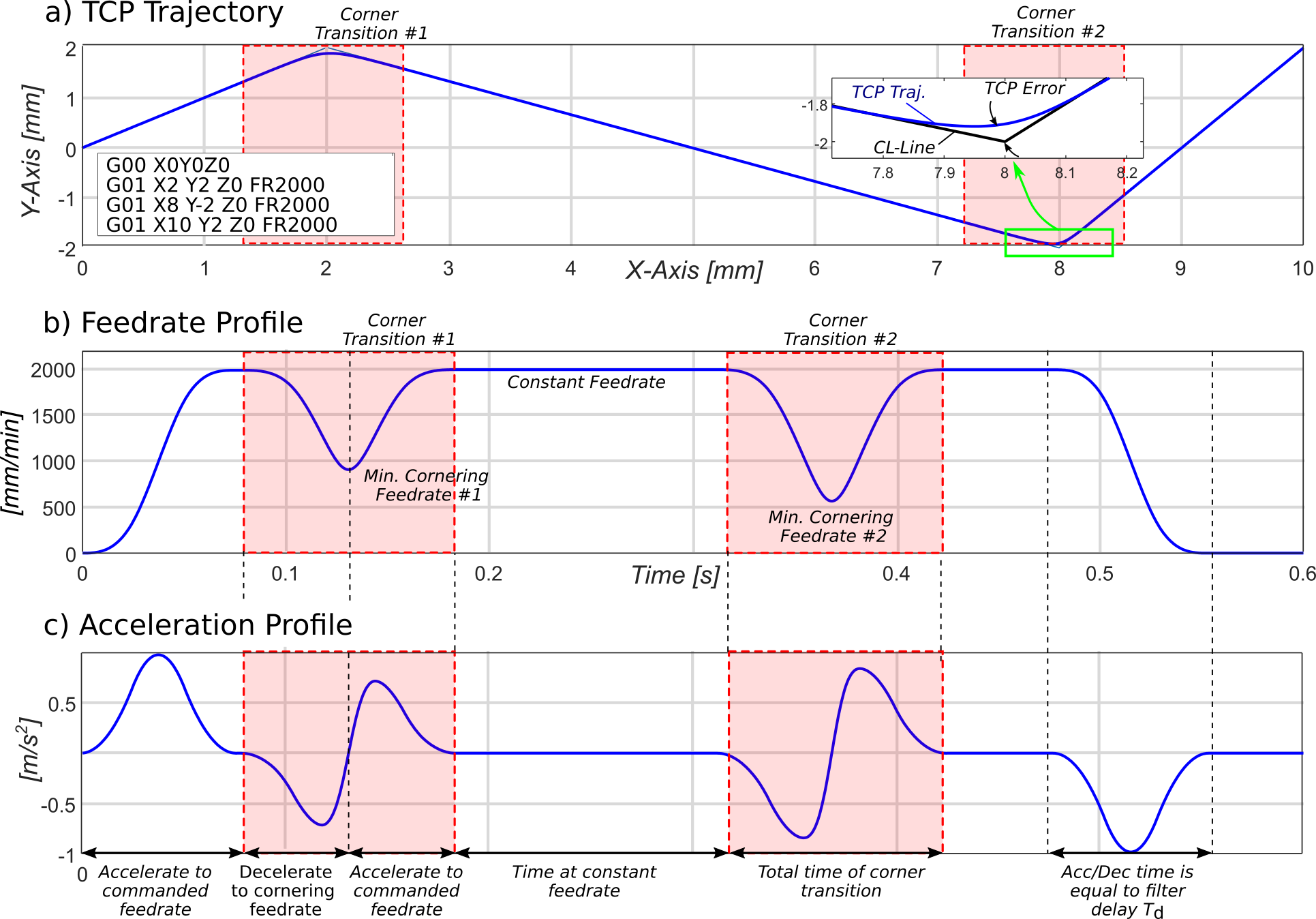}
	\caption{Typical Kinematic Profiles of an NC Program.}
	\label{Fig:Feedrate Profile}
\end{figure}

Once a part program (G-code) is deployed to a CNC machine tool, the NC unit parses the part program and interpolates the tool motion between successive cutter locations (CL). Most modern CAM systems provide tool-paths in terms of discrete CL-data and rely on linear interpolation algorithms that run in the NC units. With the introduction of cheap memory modules, long part programs do not pose a limit, and even basic circular paths are programmed with series of short linear segments \cite{Altintas2011b,Choi2007}. Therefore, modern NC systems are equipped with propriety algorithms that interpolate these lengthy series of short CL-blocks smoothly. These algorithms are called \textit{Look-ahead} or \textit{Compressor} functions and are capable of generating a non-stop motion with time optimal feed-rate profile \cite{Sencer2008} that respects kinematic limits of the machine \cite{Erkorkmaz2001a,Beudaert2012}. Prediction of a machine's actual feedrate profile requires detailed modeling of the NC system's real-time interpolation behaviour. This includes the motion transition between CL-blocks, for example a typical feedrate profile for continuous motion is shown in Fig. \ref{Fig:Feedrate Profile}. During the initial linear motion from zero to commanded feedrate the performance and behavior of the machine tool is dependent upon the acceleration and jerk constraints alone. However, as the tool approaches the end of the first CL-line (corner transition 1 in Fig.\ref{Fig:Feedrate Profile}) to change the feed direction the tool decelerates to a minimum cornering feedrate before accelerating again to the commanded feedrate. 
The reduction in feedrate in the vicinity of CL-line junction point is due to both the machine tool satisfying the tool centre point (TCP) error tolerance constraints throughout the cornering transition and the machine tool kinematic constraints. The TCP error can be seen at corner transition 2 where the TCP is maximum displacement between the CL-line and the TCP position. The TCP error constraint imposed upon the toolpath limits the maximum feedrate during cornering transitions and this significantly affects the overall machining cycle time. 

Most NC systems utilize jerk limited trajectory generation to smoothly alter feedrate and interpolate along CL-lines \cite{Erkorkmaz2001a,Jeong2005}. The generated feedrate profile is defined in the form of a cubic polynomial \cite{Altintas2011b}. Axis acceleration limits are imposed based on the torque/power capacity of the drives, and the jerk limits are set to limit unwanted vibrations during rapid feed motion \cite{Barre2005}. This general jerk-limited feedrate profile is well-known, and acceleration and jerk limits of the machine can be read from the NC system. Therefore, the use of jerk limited trajectory as a template allows prediction of feedrate kinematics of modern NC systems. Nevertheless, it can only predict machine behaviour in point-to-point (P2P) interpolation. During P2P interpolation, the tool accelerates from a full-stop to the set feedrate and decelerates again for a full-stop at the end of the CL line. Once the acceleration and jerk limits are known, the feedrate profile can be generated to predict cycle times. Past research considered modeling of NC behaviour of 3 and 5 axis machine tools for P2P trajectories \cite{Erkorkmaz2006a,Tulsyan}. 

Predicting feedrate profiles along short segmented complex toolpaths for high speed machining (HSM) is, however, a more challenging task. This is due to the fact that look-ahead modules of NC systems alter jerk limits on the fly as it blends series of CL-lines to generate a non-stop smooth continuous feed motion. Here, modeling the path blending behaviour is crucial. NC systems blend linear CL-lines together smoothly while applying geometric blending error and kinematic limit control. Machine tool literature reports that circular arcs \cite{Altintas2011b} , cubic \cite{Altintas2011b} or quintic splines \cite{Erkorkmaz2001a} can be used for such geometric path blending. There are also methods based on filtering where the discrete toolpath is blended based on low-pass filtering. Finite Impulse Response (FIR) filters are used for such purpose \cite{Tajima2018}. Such filtering based techniques are more computationally efficient and greatly favored for real-time interpolation on NC systems. For instance Heidenhain \cite{HEIDENHAIN2017,Heidenhain2011}, Mitsubishi \cite{MitsubishiElectric} and more recently Siemens \cite{SinumerikOne} NC systems utilize FIR and IIR (infinite impulse response) filters for look-ahead and non-stop smooth interpolation. Typically, users enter a blending tolerance which confines the path blending (contour) errors. Based on the blending tolerance the NC system approximates the given discrete CL-lines and plans the fastest motion with its kinematic limits. Therefore, accurate prediction of cycle times for conventional toolpaths requires modeling of NC system's non-stop interpolation behaviour along linear paths. 

This paper models the non-stop interpolation behaviour of modern NC systems and predicts feedrate profiles along HSM toolpaths by considering the real-time path blending behaviour of NC systems. Section \ref{Section:Interpolation by Filtering} briefly introduces the low-pass filtering based real-time interpolation method, which is used as a template. It is then used to predict P2P and contouring motion of NC systems in subsequent sections \ref{Section: Prediction of Kinematic Profiles} and \ref{Section:Non-Stop Interpolation}. Illustrative examples and experimental validations are provided in each section. Finally, Section \ref{Section:Validation} provides realistic cycle time, feedrate profile and cutting force prediction for complex aerospace parts.

\section{Low-pass Filtering Based Real-Time Interpolator Dynamics} \label{Section:Interpolation by Filtering}

\begin{figure}
		\centering
	\includegraphics[width=0.4\textwidth]{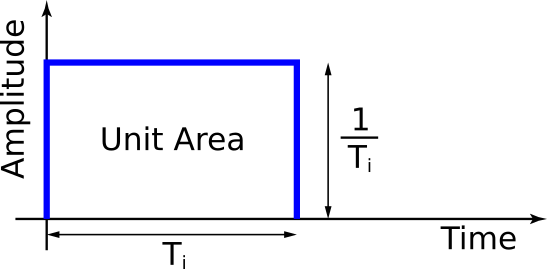}
	\caption{Impulse response of a 1st order FIR filter.}
	\label{Fig:1st Order Filter}
\end{figure}

This section models real-time interpolation behaviour of an NC system to predict the feedrate profile and overall cycle time. Most conventional NC systems utilize IIR or FIR filtering based techniques for computationally efficient real-time interpolation and feed profile planning. In this work, Finite Impulse Response (FIR) filters are used to capture the NC system's behaviour. A simple 1st order FIR filter can be expressed in the Laplace ($s$) domain by:

\begin{equation}\label{eq: M(s))}
M_{i}(s)=\frac{1}{T_{i}} \frac{1-e^{-s T_{i}}}{s},\quad i=1 \ldots n,
\end{equation}

where $s$ is a complex number,  $T_{i}$ is the time constant of the $i^{th}$ filter. The impulse response is depicted in Fig. \ref{Fig:1st Order Filter}. As seen in \eqref{eq: M(s))}, the filter contains an integrator, which acts to smooth  the input signal. These two features of 1st order FIR filters are appealing from a NC system perspective, since G-codes (represented by rectangular velocity pulses) can be convolved through a series of such filters to generate smooth velocity profiles. Since the area underneath the rectangular impulse response is unitary, the area underneath the original input is not altered. 
 \cite{Tajima2018,HEIDENHAIN2017,MitsubishiElectric,SinumerikOne,Tajima2019,Tajima2020}.

\begin{figure}[ht]
	\centering
	\includegraphics[width=\textwidth]{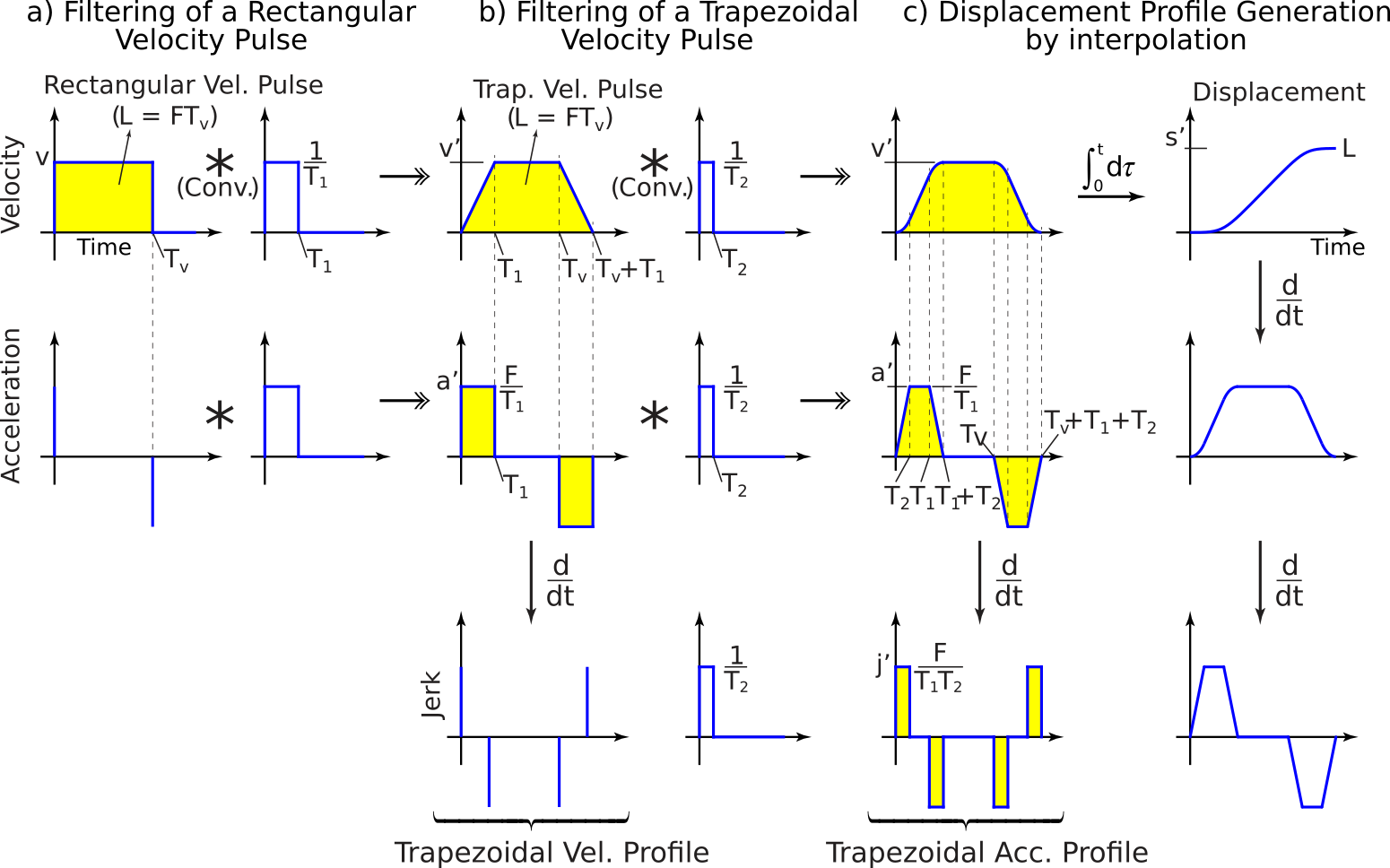}
	\caption{Smooth trajectory generation by low order FIR filtering \cite{Tajima2018}}
	\label{Fig:Low Order Traj Gen}
\end{figure}

Fig.\ref{Fig:Low Order Traj Gen} illustrates this filtering based interpolation procedure. As shown, consider a G-code for a total displacement command of $L$ at a feedrate of $F$. It is represented by a velocity pulse with an amplitude of $F$ and duration of $T_{v}$ hence $L = FT_{v}$. Subsequent convolution of the velocity pulse with the FIR filter yields the higher order velocity response. Using 2-FIR filters in series generates reference trajectories with piece-wise constant jerk profiles and using three FIR filters in series further smooths the reference velocity making them snap limited. Although jerk-limited trajectories are most common in high speed machinery, snap limited trajectories are tuned for ultra-precision machines \cite{Lambrechts2005} to further mitigate the effect of unwanted vibrations.

The duration of the original velocity pulse $T_{v}$ and the time constants of the filters $T_{1}$, $T_{2}$ and $T_{3}$ determine the velocity and acceleration profiles, which can be derived analytically by evaluating the convolution integral between the input velocity pulse and the rectangular impulse response of the filter as follows:

\begin{equation}
\begin{aligned}
v^{\prime}(t)=& v(t) * m(t) \\
&=\frac{1}{T_{1}} \int_{0}^{t}\left(\left[v(\tau)-v\left(\tau-T_{v}\right)\right]\left[u(t-\tau)-u\left(t-T_{1}-\tau\right)\right]\right) d \tau \\
=&\frac{1}{T_{1}}\left[\begin{array}{l}
\int_{0}^{t} v(\tau) u(t-\tau) d \tau-\int_{0}^{t} v(\tau) u\left(t-T_{1}-\tau\right) d \tau \\
-\int_{0}^{t} v\left(\tau-T_{v}\right) u(t-\tau) d \tau+\int_{0}^{t} v\left(\tau-T_{v}\right) u\left(t-T_{1}-\tau\right) d \tau
\end{array}\right]
\end{aligned}
\end{equation}

where $v(t)$,$v'(t)$ and $m(t)$ represent the velocity pulse, interpolated velocity signal and  the impulse response of the FIR filter (Eq.\eqref{eq: M(s))} respectively.

For $T_{2}>T_{1}>T_{v}$ the velocity and acceleration profiles become:

\begin{equation}\label{eq:FIR2 Vel}
v^{\prime}(t)=\left\{\begin{array}{ll}
\frac{1}{2} \frac{F}{T_{1} T_{2}} t^{2} & 0 \leq t<T_{2} \\
\frac{1}{2} \frac{F T_{2}}{T_{1}}+\frac{F}{T_{1}}\left(t-T_{2}\right) & T_{2} \leq t<T_{1} \\
F-\frac{1}{2} \frac{F}{T_{1} T_{2}}\left(\left(T_{1}+T_{2}\right)-t\right)^{2} & T_{1} \leq t<T_{1}+T_{2} \\
F & T_{1}+T_{2} \leq t<T_{v} \\
F-\frac{1}{2} \frac{F}{T_{1} T_{2}}\left(t-T_{v}\right)^{2} & T_{v} \leq t<T_{v}+T_{2} \\
F-\frac{1}{2} \frac{F T_{2}}{T_{1}}-\frac{F}{T_{1}}\left(t-\left(T_{v}+T_{2}\right)\right) & T_{v}+T_{2} \leq t<T_{v}+T_{1} \\
\frac{1}{2} \frac{F}{T_{1} T_{2}}\left(\left(T_{v}+T_{1}+T_{2}\right)-t\right)^{2} & T_{v}+T_{1} \leq t<T_{v}+T_{1}+T_{2} \\
0 & T_{v}+T_{1}+T_{2} \leq t
\end{array}\right.\end{equation}

\begin{equation}\label{eq:FIR2 Acc}
a^{\prime}(t)=\left\{\begin{array}{ll}
\frac{F}{T_{1} T_{2}} t & 0 \leq t<T_{2} \\
\frac{F}{T_{1}} & T_{2} \leq t<T_{1} \\
\frac{F}{T_{1}}-\frac{F}{T_{1} T_{2}}\left(t-T_{1}\right) & T_{1} \leq t<T_{1}+T_{2} \\
0 & T_{1}+T_{2} \leq t<T_{v} \\
-\frac{F}{T_{1} T_{2}}\left(t-T_{v}\right) & T_{v} \leq t<T_{v}+T_{2} \\
-\frac{F}{T_{1}} & T_{v}+T_{2} \leq t<T_{v}+T_{1} \\
-\frac{F}{T_{1}}+\frac{F}{T_{1} T_{2}}\left(t-\left(T_{v}+T_{1}\right)\right) & T_{v}+T_{1} \leq t<T_{v}+T_{1}+T_{2} \\
0 & T_{v}+T_{1}+T_{2} \leq t
\end{array}\right.\end{equation}

It is shown in Fig.\ref{Fig:Low Order Traj Gen}a that when a square velocity pulse is convolved with a first order FIR filter the result is a trapezoidal velocity profile. In this case the time constant of the FIR filter $T_{1}$ is smaller than the length of the
velocity pulse $T_{v}$, ($T_{1}<T_{v}$). It can be seen that the resultant length of the velocity profile is extended to $T_{v} + T_{1}$, the length of the filtered velocity profile is elongated by the length of the FIR filter. The figure also shows the profile is segmented into 3 main kinematic sections. The figure shows that when $T_{1} < T_{v}$
the commanded feedrate $F$ is reached and maintained
for the cruise duration $T_{v}-T_{1}$. The filtered acceleration profile shows the peak acceleration is $V/T_{1}$, demonstrating
the FIR filter time constant governs the acceleration properties of the filtered kinematic profile. When $T_{1}>T_{v}$ then the commanded velocity cannot be reached and the maximum velocity is determined by $L/T_{1}$.

Cascading FIR filters can increase the order (smoothness) of the kinematic profile as is shown in Fig.\ref{Fig:Low Order Traj Gen}b. Using an FIR filter with a time constant of $T_{2}$, where $T_{2}<T_{1}$, the trapezoidal velocity profile is filtered
once more. The resultant motion profile is constructed from 7 segments in which the time of each segment is determined from $T_{1}$, $T_{2}$ and $T_{v}$. It can be seen that the total duration is $T_{1}+T_{2}+T_{v}$. Due to the linearity of the convolution operation \cite{Smith2003} the final kinematic profiles are independent of the order of the filtering, it is the relationship between the magnitudes of the time constants that determines the maximum feedrate and acceleration \cite{Tajima2018}. The analytical expressions of velocity and acceleration for each segment are shown in equations \eqref{eq:FIR2 Vel} and \eqref{eq:FIR2 Acc} respectively.

\begin{figure}[ht]
    \centering
    \includegraphics[width=0.5\textwidth]{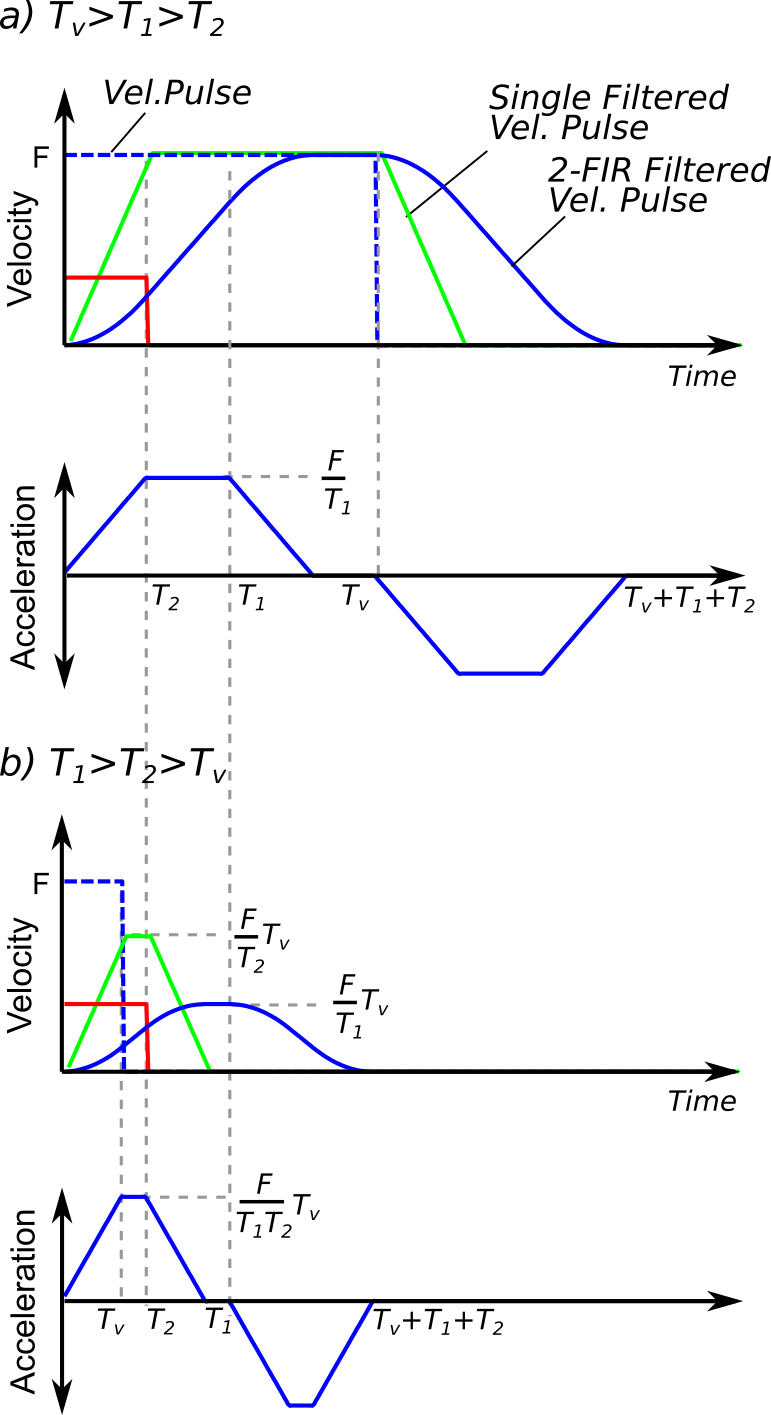}
    \caption{Trapezoidal acceleration profile generated by 2 FIR filters with different time constants.}
    \label{fig:Filters_Different}
\end{figure}

Figs. \ref{fig:Filters_Different}a and \ref{fig:Filters_Different}b show example profiles for different velocity pulse and filter delay parameters.
As shown, the maximum acceleration and jerk values of the profiles are determined by the time constants of the FIR filter. In Fig.\ref{fig:Filters_Different}a the commanded feedrate is reached on the precondition that $T_{v}>T_{1}+T_{2}$ and $T_{1}>T_{2}$ and in this case the peak acceleration can be determined from the longest FIR filter time constant $V/T_{1}$. However, as shown in Fig.\ref{fig:Filters_Different}b where $T_{v}$ is smaller than the smallest filter delay $T_{1}>T_{2}>T_{v}$, then the commanded feedrate $F$ is not reached. The maximum feedrate and acceleration is determined by the commanded feedrate and the relationship of the filter time constants $T_{1}$,$T_{2}$ and the velocity pulse $T_{v}$. The maximum feedrate and acceleration are calculated from $FT_{v}/T_{1}$ and $FT_{v}/T_{1}T_{2}$ respectively. Note the maximum feedrate is limited by the longest time constant and velocity pulse length $T_{v}$, whereas the acceleration is constrained by both time constants.

\begin{figure}[ht]
		\centering
	\includegraphics[width=0.5\textwidth]{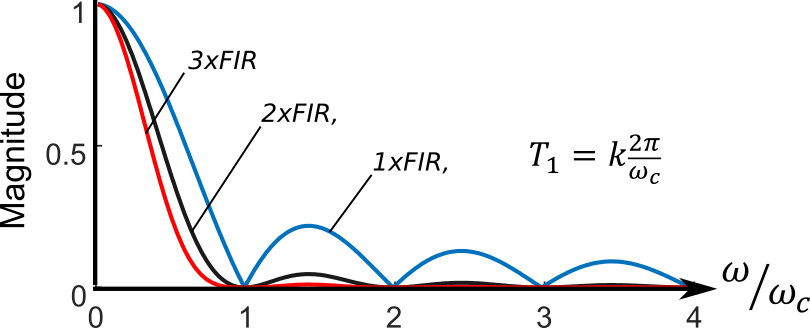}
	\caption{Magnitude of the frequency response of multiple FIR filters}
	\label{Fig:Frequency Response}
\end{figure}

Time constants of the filters are typically selected to mitigate structural vibrations on the machine \cite{Biagiotti2012}. Matching the time constant with the vibration period of the lightly damped modes helps avoid exciting them during rapid acceleration. 
Another way to specify the time constants is to set them equal, $T_{1}=T_{2}$. In this special case, the FIR filter acts as a pure low pass filter with a roll-over frequency of $\omega_{c}=\frac{2\pi}{T_{1}}$. Fig.\ref{Fig:Frequency Response} shows the attenuation in the frequency response  for multiple FIR filters with matching time constants. The time constant, when set low enough, helps prevent the excitation of any higher frequency vibrations during rapid accelerations. This simpler method compared to tuning individual filters provides a convenient method of vibration suppression during high feedrates. 

For interpolation using 2-FIR filters with matching time constants, the transfer function of the resulting FIR filter is:

\begin{equation}\label{eq:M_2FIR(2)}
M_{2FIR}(s)= \left(\frac{1}{T_{1}} \frac{1-e^{-s T_{1}}}{s}\right)\left(\frac{1}{T_{1}} \frac{1-e^{-s T_{1}}}{s}\right)
\end{equation}

and the resulting velocity profile when a rectangular feed pulse $v(t)$ is filtered $T_{1}=T_{2}<T_{v}$ becomes:

\begin{equation}
    v^{\prime}(t)= v(t) * m_{2FIR}(t) 
\end{equation}

\begin{equation}\label{eq:FIR 2 T1 Vel}
v^{\prime}(t)=\left\{\begin{array}{ll}
\frac{F}{2T_{1}^{2}} t^2 & 0 \leq t<T_{1} \\
\frac{F}{2T_{1}}\left(-t^2+4T_{1}t-2T_{1}^2\right) & T_{1} \leq t<2T_{1} \\
F & 2T_{1} \leq t<T_{v} \\
\frac{F}{2T_{1}}\left(-t^2+2T_{v}t-T_{v}^2+2T_{1}^2\right) & T_{v} \leq t<T_{v} + T_{1} \\
\frac{F}{2T_{1}^2}\left(t^2-2T_{v}t-4T_{1}t + \left(T_{v}+2T_{1}\right)^2\right) & T_{v}+T_{1} \leq t<T_{v}+2T_{1} 
\end{array}\right.\end{equation}

The corresponding acceleration and jerk responses are : 

\begin{equation}\label{eq:FIR 2 T1 Acc}
a^{\prime}(t)=\left\{\begin{array}{ll}
\frac{F}{T_{1}^{2}} t & 0 \leq t<T_{1} \\
\frac{F}{T_{1}^2}\left(-t+2T_{1}\right) & T_{1} \leq t<2T_{1} \\
0 & 2T_{1} \leq t<T_{v} \\
\frac{F}{T_{1}^2}\left(-t+T_{v}\right) & T_{v} \leq t<T_{v} + T_{1} \\
\frac{F}{T_{1}^2}\left(t-T_{v}-2T_{1}\right) & T_{v}+T_{1} \leq t<T_{v}+2T_{1} 
\end{array}\right.\end{equation}

\begin{equation}\label{eq:FIR 2 T1 Jerk}
j^{\prime}(t)=\left\{\begin{array}{ll}
\frac{F}{T_{1}^2} & 0 \leq t<T_{1} \\
-\frac{F}{T_{1}^2} & T_{1} \leq t<2T_{1} \\
0 & 2T_{1} \leq t<T_{v} \\
-\frac{F}{T_{1}^2} & T_{v} \leq t<T_{v} + T_{1} \\
\frac{F}{T_{1}^2} & T_{v}+T_{1} \leq t<T_{v}+2T_{1} 
\end{array}\right.\end{equation}

\begin{figure}
    \centering[ht]
	\includegraphics[width=0.5\textwidth]{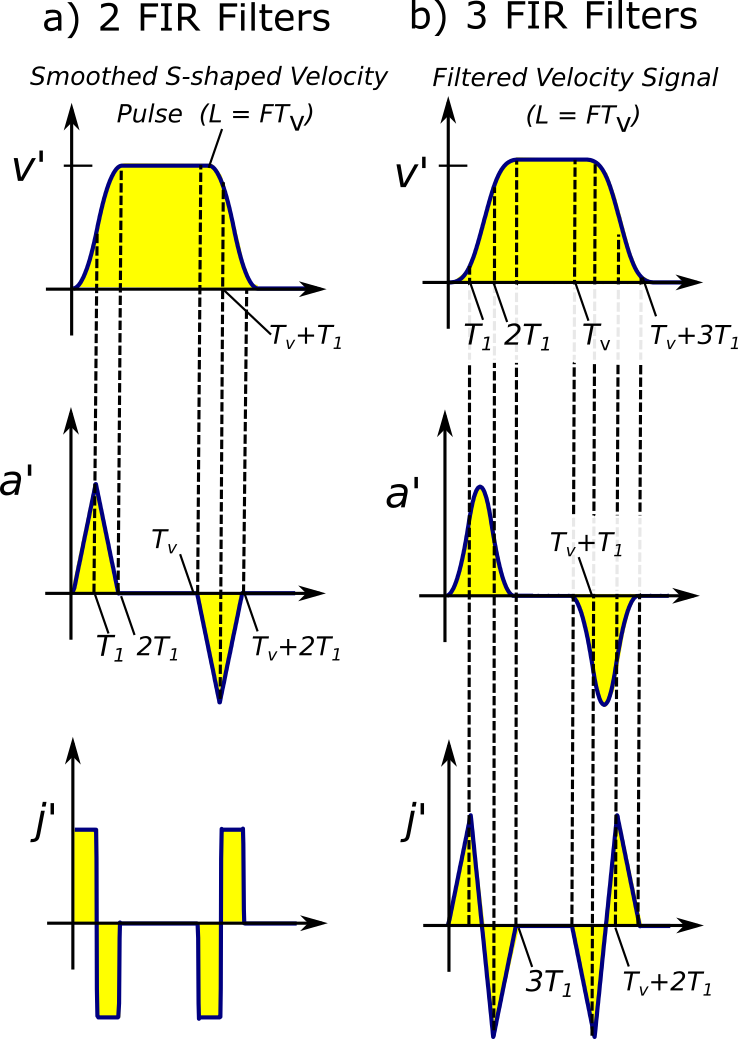}
	\caption{Velocity, acceleration and jerk profiles for the 2 filter case ($T_{1}=T_{2}$) and 3 filter case ($T_{1}=T_{2}=T_{3}$)}
	\label{Fig:Filter 3 Low Order}
\end{figure}

As demonstrated, when a square velocity pulse of magnitude $F$ and length $T_{v}$ is convolved with a first order FIR filter with time constant $T_{1}$ the result is a trapezoidal velocity profile with constant acceleration of magnitude $F/T_{1}$ (Fig. \ref{Fig:Filter 3 Low Order}a). The total length of the kinematic profiles are extended by the filter time constant $T_{1}$ to $T_{v}+T_{1}$. When the trapezoidal velocity profile is convolved with a second first order FIR filter with a matching time constant $T_{1}=T_{2}$ the smoothness (order) of the velocity profile is increased from $C^1$ to $C^2$, where $C^{n}$ is the space of n\textsuperscript{th} order continuously differentiable functions, as shown in equations \eqref{eq:FIR 2 T1 Vel} and Fig. \ref{Fig:Filter 3 Low Order}b. However, using the matching time constant $T_{1}=T_{2}$, results in five sections in the kinematic profile and not seven as for the case for two different time constants where $T_{1}\neq T_{2}$. The resulting acceleration profile is triangular around $T_{1}$ and $T_{v}+T_{1}$ with peak magnitudes $F/T_{1}$ and lengths of $2T_{1}$; the now jerk limited profile has peak magnitudes of $F/T_{1}^2$. The total length of the kinematic profiles is extended to $T_{v}+2T_{1}$. The relationship between $T_{1}$ and $T_{v}$ determines the kinematic constraints as for the different filter cases. For completeness, the authors include the example profiles for the matching filter case in appendix \ref{Appendix Filter Matching} for different velocity pulse and filter delay parameters.

Convolving the velocity profile with a third first order FIR filter with the same time constant $T_{1}=T_{2}=T_{3}$ results in a $C^3$ velocity profile, $C^2$ acceleration profile and $C^1$ jerk profile. The velocity, acceleration and jerk equations for the 3-FIR case is shown in appendix \ref{Appendix 3 Filter Equations}. The smooth acceleration profile has a peak magnitude of $3F/4T_{1}$ at times $1.5T_{1}$ and $T_{v}+1.5T_{1}$ and the jerk profile has peak magnitudes of $F/T_{1}^2$. The overall length of the kinematic profiles have been extended from the original square velocity pulse length $T_{v}$ to $T_{v}+3T_{1}$. The total filter delay when using 3-FIR filters with matching time constants $T_{1}$ is therefore $3T_{1}$.

It can be shown that a high order FIR filter can be accurately modelled and implemented with using only 3 first order FIR filters. The benefit of using 3 or more first order FIR filters with the same time constant is that the filter response approaches that from a Gaussian filter. The Gaussian response has no overshoot whilst minimising the acceleration and deceleration time periods which makes it the ideal time domain filter for interpolating kinematic profiles \cite{Blinchikoff}. The ability to approximate the Gaussian filter with 3 FIR filters with the same time constant simplifies the design and selection of the filter to a single design parameter $T_{1}$. For both the 2 and 3 FIR filter cases, $T_{1}$ can be analytically calculated from the maximum permissible jerk $J_{max}$ as follows: 

\begin{equation}\label{eq:Jerk Max}
J_{max} = \frac{\Delta F}{T_{1}^2},\longrightarrow T_{1} = \sqrt{\frac{\Delta F}{J_{max}}} 
\end{equation}

\subsection{Identification of Real-Time Interpolator Dynamics of an NC system}

\begin{figure}
    \centering
    \includegraphics[width=\textwidth]{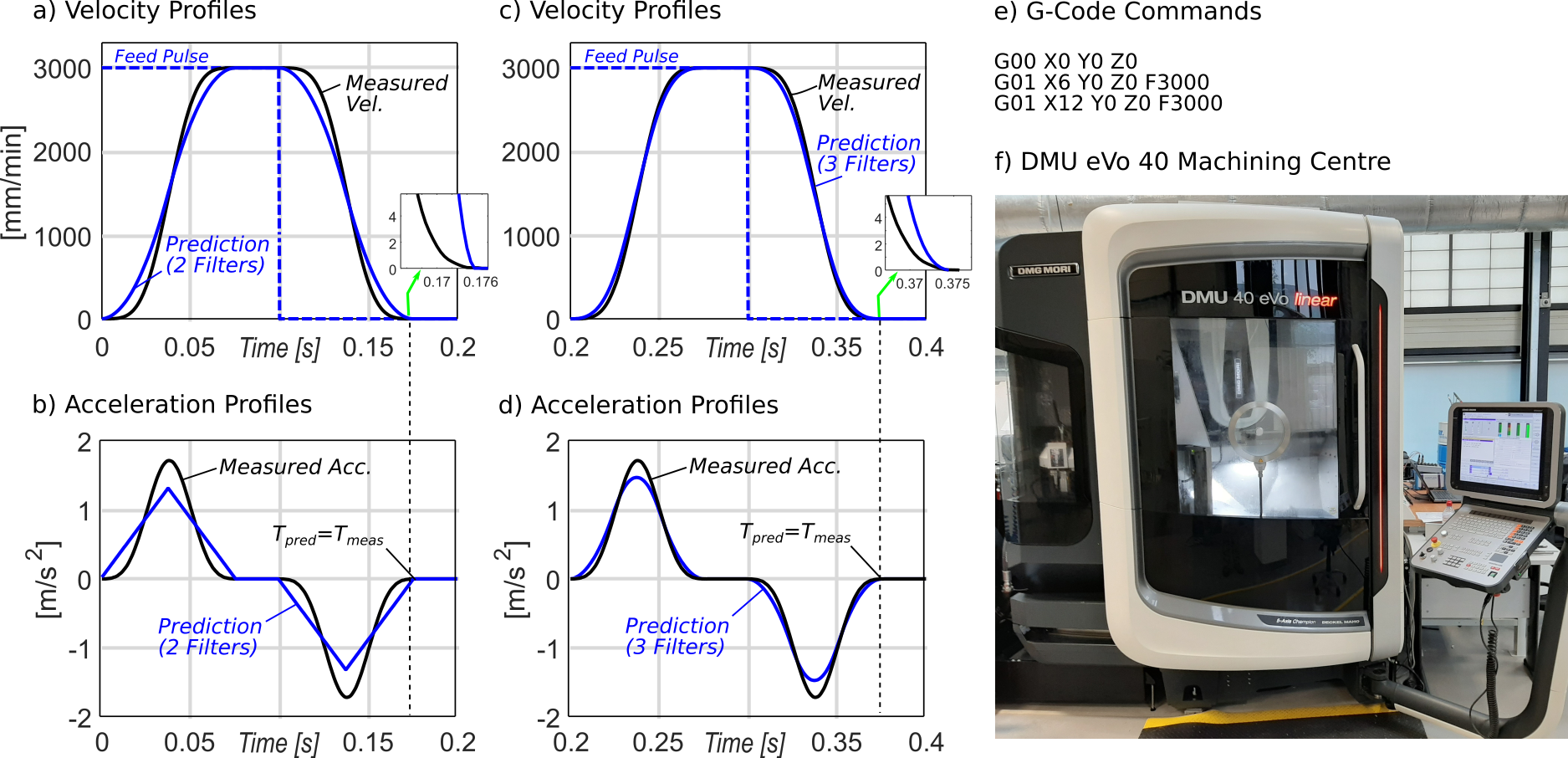}
    \caption{Measured and predicted velocity and acceleration profiles for 2-FIR (a) and (b) and 3-FIR (c) and (d) P2P motion interpolation}
    \label{fig:DMU 340}
\end{figure}

The previous section presented the filtering based real-time trajectory generation. In this section it is shown how the interpolator response of a machine tool can be modelled via the identification of the filter time-constants. A case study was conducted on the DMG Mori eVo40 machine tool shown in Fig.\ref{fig:DMU 340}. The machine is commanded by a single G-code to move 6 mm at a speed of 3000 mm/min, and the interpolated reference motion profile is recorded on the NC system directly at a sampling time of $T_{s}=0.009s$. Figs \ref{fig:DMU 340}a to \ref{fig:DMU 340}b, show the recorded kinematic profiles. The machine is set to undergo a simple point-to-point (P2P) motion and therefore the tool comes to a full stop before moving to the next commanded position. As shown for the measured system, the NC system generates smooth velocity and acceleration profiles. The acceleration profile mimics a smooth 'bell-shaped' profile. Overall, acceleration, and deceleration duration are measured to be $T_{acc}=T_{dec} = 0.0765$ sec. The cruise velocity portion is roughly measured to be 0.023 sec. 
In order to simulate the feed profile, a series of 2 and 3-FIR filters are used. For the 2-FIR case the time constant is selected as $T_{1}=\frac{T_{acc}}{2}$ and for the 3-FIR case it is set to $T_{1} = \frac{T_{acc}}{3}$. The predicted velocity and acceleration profiles for the 2-FIR case are shown in Figs \ref{fig:DMU 340}a and \ref{fig:DMU 340}b respectively. The time of the measured displacement is equal to the time of the predicted displacement. The difference between the velocity profiles is due to the acceleration. The 2-FIR case exhibits the triangular acceleration profile compared to the smooth measured response. The maximum acceleration for the 2-FIR case is constrained and less than the measured response. 

In order to compare the different filter cases the machine is commanded to move along the same G-code, and the proposed interpolator model for the 3-FIR case is used. As shown in Fig.\ref{fig:DMU 340}c the velocity profiles for the 3-FIR case closely resembles the measured velocity profile and the total time of the measured displacement matches the total time for the simulated displacement. The simulated acceleration profile is smooth and the maximum acceleration is higher than for the 2-FIR case but still lower than the measured response. Increasing the order of the simulated system would allow the maximum acceleration to approach the measured response. In general, by increasing the order of the FIR filter, the predicted acceleration profile of the filtered pulse approaches the acceleration profile of the measured response and results in a simulated velocity profile which closely resembles the dynamics of the machine interpolator.

The filter delay is calculated from the jerk \eqref{eq:Jerk Max} and the duration of the acceleration phase in each case is equal to the total filter delay. The time constant (filter delay) can be analytically calculated from machine tools' specifications ($J_{max}$) and therefore kinematic profiles can be generated using FIR filters without the requirement for parameter identification through system testing.

In this section it has been shown that the dynamics of an NC interpolator are increasingly well-approximated by the series combination of identical first-order FIR filters. In addition, the relationship between the parameters of these first-order filters and the resulting interpolator response have been derived.

\section{Multi-Axis P2P Motion Generation}\label{Section: Prediction of Kinematic Profiles}

FIR filtering based interpolation of single axis motion was presented in the previous sections. Extending the method to P2P multi-axis linear motion this section describes the process to interpolate kinematic profiles between two points using high order FIR filters.

\begin{figure}[ht]
	\centering
	\includegraphics[width=0.5\textwidth]{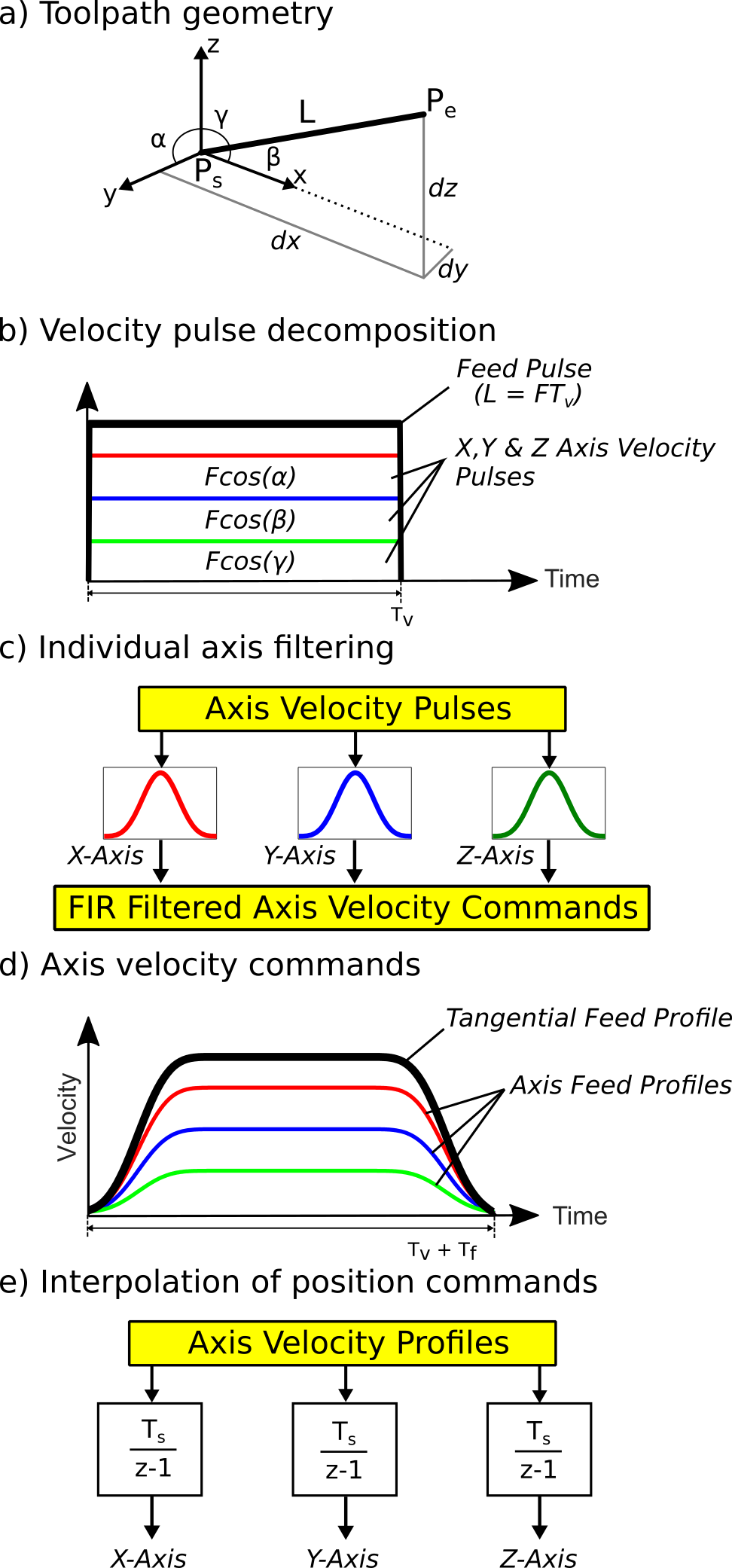}
	\caption{Multi-axis interpolation based on high order FIR filtering.}
	\label{Fig:Multi-Axis High Order FIR}
\end{figure}

% \subsection{} \label{Section:P2P}
The start and end positions of a linear G01 command in 3 axes can be represented by $\mathbf{P}_{\mathrm{s}}=\left[P_{s, x}, P_{s, y}, P_{s, z}\right]^{T}$
and $\mathbf{P}_{\mathrm{e}}=\left[P_{e, x}, P_{e, y}, P_{e, z}\right]^{T}$, respectively as shown in Fig. \ref{Fig:Multi-Axis High Order FIR}a.

The tool displacement $L$ is calculated by taking the
Euclidean norm of the vector between the two commanded positions, $L=\left\|\mathbf{P}_{e}-\mathbf{P}_{s}\right\|_{2}$.The velocity pulses of each axis $\left(v_{x}, v_{y}, v_{z}\right)$ are calculated by multiplying the feed pulse $v(t)$ by the unit
velocity vector $\mathbf{u}=(\mathbf{P_{e}}-\mathbf{P_{s}}) /\|\mathbf{P_{e}}-\mathbf{P_{s}}\|_{2}$.

\begin{equation}\frac{d \mathbf{P}(\mathbf{t})}{d t}=\dot{\mathbf{P}}(\mathbf{t})=v(t) \mathbf{u}=\left[\begin{array}{c}
v_{x}(t) \\
v_{y}(t) \\
v_{z}(t)
\end{array}\right]\end{equation}

where $\dot{\mathbf{P}}(\mathbf{t})$ represents the first time derivative of the P2P displacement (Fig. \ref{Fig:Multi-Axis High Order FIR}b).

In order to generate (and interpolate) the reference velocity commands ($v^{\prime}_{x}$,$v^{\prime}_{y}$,$v^{\prime}_{z}$), the individual axis velocity pulses
($v_{x}$, $v_{y}$, $v_{z})$
are convolved with the FIR filter (Figs. \ref{Fig:Multi-Axis High Order FIR}c and \ref{Fig:Multi-Axis High Order FIR}d):

\begin{equation}\label{eq:Interpolate}
\frac{d \mathbf{P}^{\prime}(t)}{d t}=\dot{\mathbf{P}}^{\prime}(t)=\left[\begin{array}{c}
v_{x}^{\prime}(t) \\
v_{y}^{\prime}(t) \\
v_{z}^{\prime}(t)
\end{array}\right]=\dot{\mathbf{P}}(t) * m(t)\end{equation}

Finally, the filtered position commands are generated by integrating the filtered axis velocity commands:

\begin{equation}\label{eq: position integration}
\mathbf{P}^{\prime}(t)=\left[\begin{array}{c}
p_{x}^{\prime}(t) \\
p_{y}^{\prime}(t) \\
p_{z}^{\prime}(t)
\end{array}\right]=\int_{0}^{t} \left[\begin{array}{c}
v^{\prime}_{x}(t) \\
v^{\prime}_{y}(t) \\
v^{\prime}_{z}(t)
\end{array}\right] \, d \tau \ \end{equation}

\section{Prediction of Interpolator Behaviour during Non-stop High Speed Motion}\label{Section:Non-Stop Interpolation}

The previous section showed that P2P linear interpolation behaviour of an NC system can be modelled by velocity pulses low pass filtered by a series of first order FIR filters. The only required parameter to predict the machine's feed profile and accurately estimate the resulting cycle time is the time constant, i.e. total delay of the FIR filter. As shown, the filter time delay can be calculated from the maximum permissible jerk \eqref{eq:Jerk Max} and commanded feedrate.
This section focuses on accurate prediction of interpolator behaviour during non-stop contouring motion, which is the most commonly used interpolation technique for high speed machining (HSM).

\subsection{Modeling of Non-stop Interpolation Behaviour} \label{Section:Velocity Blending}
Typical high speed machining toolpaths found in die and mould manufacturing or in aerospace industry consist of series of short segmented toolpaths \cite{10.1007/978-1-84996-432-6_60}. When interpolated in HSM mode, the NC interpolator does not undergo a full-stop at the end of each CL line. Instead, the CL lines are blended together for a non-stop smooth motion interpolation where machining feedrate is reduced to a cornering speed $V_{c}$ around junction points of the CL-blocks (See Fig.\ref{Fig:Feedrate Profile}). The prediction of $V_{c}$ is crucial to accurately capture the actual feedrate profile and estimate the resultant cycle time. Several constraints affect the cornering speed ($V_{c}$) and overall acceleration profile around the CL data points. Firstly, $V_{c}$ is controlled by the blending (cornering) tolerance \cite{Erkorkmaz2006a}. Typically, lower blending tolerance delivers more accurate motion but generates slower feed profiles. In contrary, a larger tolerance value allows faster speeds and shorter overall cycle time. The relationship between the blending tolerance and the feed drop around the corner must be captured. Secondly, the deceleration/acceleration profile and the transition duration from the programmed feedrate ($F$) to the cornering speed ($V_{c}$) are dictated by acceleration and jerk limits of the machine. Both of these key characteristics must be modelled to accurately predict the varying feedrate profile along HSM tool-paths. 

\begin{figure}
    \centering
    \includegraphics[width=0.5\textwidth]{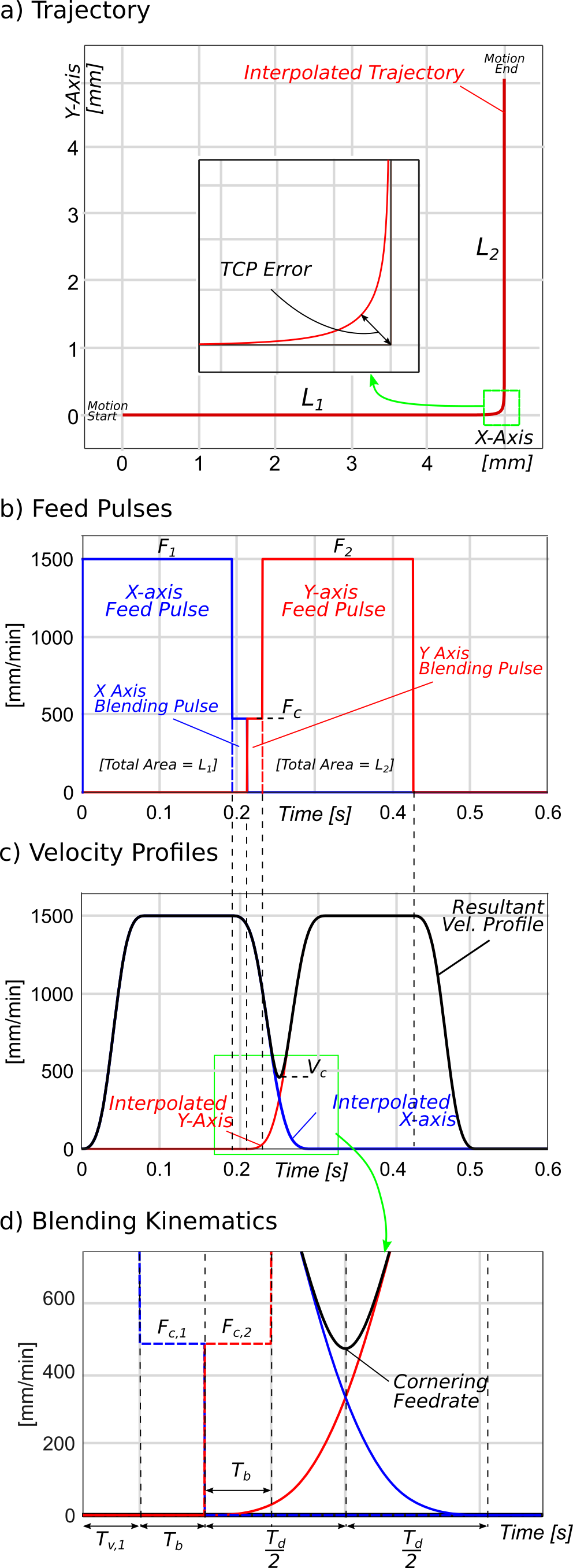}
    \caption{FIR based interpolation of a right angled toolpath with a constant feedrate}
    \label{fig:Corner Traj and Profile}
\end{figure}

In an effort to accurately model the interpolator behaviour, the feed pulse distribution shown in Fig. \ref{fig:Corner Traj and Profile}b is proposed in this manuscript. Notice that the feed pulse profile is different from the case used for the P2P motion. Feed pulses of each CL block are commanded back-to-back with no dwell time in between. In other words, they are constructed as a continuous pulse stream. The duration of the feed pulse is $T_{v}$. Notice that the feed pulse does not have a constant amplitude of $F$. Instead, around CL block junctions the feed command value is dropped down to $F_{c}$. Such small feed pulse is added to model the blending kinematics, commanding the feedrate to drop down to a cornering feed of $F_{c}$. The duration of the cornering feed pulse is set to $T_{b}$, which controls how long the deceleration and acceleration last around the blend.

When the feed pulse profile is interpolated with a FIR filter the resulting velocity profiles are smooth velocity profiles that better approximate the actual velocity profiles of the machining interpolator. Fig.\ref{fig:Corner Traj and Profile}a and Fig.\ref{fig:Corner Traj and Profile}c show the toolpath and the corresponding interpolated X-axis and Y-axis velocity profiles respectively. The total length of the velocity profiles is equal to the sum of the pulse lengths plus the filter delay $T_{d}$. Fig.\ref{fig:Corner Traj and Profile}d shows the cornering feedrate $V_{c}$ of the resultant velocity profile is equal to the commanded blending pulse feedrate $F_{c}$ and this occurs at half the filter delay $T_{d}/2$ from the start of the Y-axis profile. 

The cornering feedrate is controlled by setting the blending velocity pulse $F_{c}$ equal to the desired cornering tangential velocity $V_{c}$ and setting the acceleration and deceleration time for the interpolated feed profile equal to the time required to reduce from $F$ to $F_{c}$. A scaling factor is applied to $F$ to represent $F_{c}$ as a function of commanded feedrate $F$:

\begin{equation}\label{eq:alpha}
    F_{c} = F\alpha = V_{c}
\end{equation}

where $V_{c}$ is the resultant 3-axis TCP velocity defined as

\begin{equation}\label{eq:vmin}
    V_{c} = \sqrt{v_{x}^{\prime2}+v_{y}^{\prime2}+v_{z}^{\prime2}}
\end{equation}

and $v_{x}^{\prime},v_{y}^{\prime}$ and $v_{z}^{\prime}$ represent the interpolated axis velocities at the minimum corning feedrate.

The total acceleration and deceleration time of the interpolated feed profile to reach $F\alpha$ from $F$ is represented by $T_{b}$, it is a function of the filter delay $T_{d}$, and it can be calculated as:

\begin{equation} \label{Eq: Tb Equation}
    T_{b} = \frac{1}{2}T_{d}\left(1-\alpha\right)
\end{equation}

The final component to the pulse train is determining the main velocity pulse lengths $T_{v}$. In section \ref{Section:Interpolation by Filtering} the length of the velocity pulse  $T_{v}$ was calculated from $L/F$, however, with the introduction of the blending pulses, $T_{v}$ must be modified in order to preserve the total area of the pulses and hence the TCP displacement.

The commanded TCP displacement is calculated from the total area of the velocity pulse and the blending pulse, this can be seen in Fig.\ref{fig:Corner Traj and Profile}b  where the total area within the X-axis and Y-axis pulses is equal to $L_{1}$ and $L_{2}$ respectively. 

For a single axis displacement $L$ the pulse areas comprise of the main pulse (calculated as $FT_{v}$) and the blending pulse (calculated as $F_{c}T_{b}$):

\begin{equation} \label{eq:L}
L=F T_{v}+F_{c} T_{b} 
\end{equation}

Rearranging equation \eqref{eq:L} and incorporating equation \eqref{eq:alpha} yields the modified value of $T_{v}$ as:   

\begin{equation}\label{eq:Tv - single sideed}
T_{v}=\frac{L}{F} - \alpha T_{b}\end{equation}

Equation \ref{eq:Tv - single sideed} holds for velocity commands with a single blending pulse, this is the case for the initial and final CL lines in a part program which start and end at zero feedrate (full stop). The remaining displacements in a part program are continuous and therefore the commands consist of a velocity pulse with a blending pulse either side as shown in Fig.\ref{Fig:FIR Process}. Therefore each cornering blend consists of two back to back blending pulses. 

For the entire pulse train, each G01 command or CL-line can be represented by an index k with k=1 corresponding to the initial command in the part program. The associated feedrate commands in the part program are hence denoted~$F(k)$. Therefore, for the main commands in a part program the modified value of $T_{v}$ is calculated as:

\begin{equation}\label{eq:Tv main}
T_{v}(k)=\frac{L(k)}{F(k)}-\alpha(k) T_{b}(k) -\alpha(k+1) T_{b}(k+1)\end{equation}

For constant feedrate the adjoining blending pulses are symmetric. This leads to symmetrical interpolated velocity profiles and results in symmetrical displacement profiles, translating to the same toolpath trajectory for both forward and backward passes resulting in a more accurate finish.

\subsection{Filtered Signal Generation}\label{Section:Signal Generation}

\begin{figure}[]
		\centering
	\includegraphics[width=0.5\textwidth]{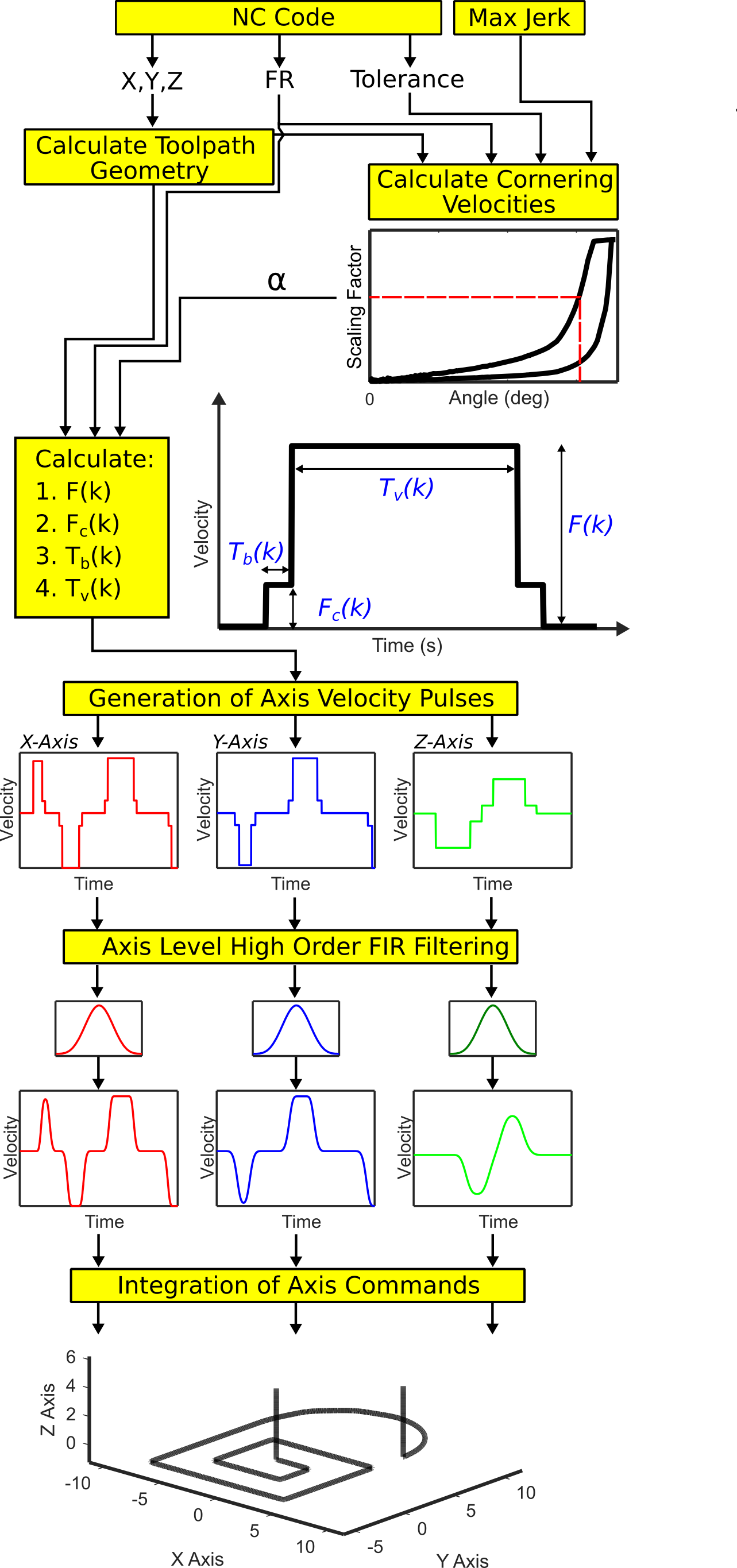}
	\caption{Non-stop interpolation of kinematic profiles using high order FIR filtering}
	\label{Fig:FIR Process}
\end{figure}

The composition of the velocity pulses and filtered kinematic profiles was shown in the previous section. In practise, the strategy for interpolation of multi-segmented NC tool-paths using high order FIR filtering, as shown in Fig. \ref{Fig:FIR Process}, is as follows:

\begin{enumerate}
	\item Read NC code and parse commanded X,Y, Z positions and feedrate commands from individual G01 commands and extract defined tolerance setting.
	\item Calculate toolpath geometry (cornering angles $\theta_{TCP}$) and unit velocity vectors for each G01 command.
	\item Calculate cornering feedrates where $V_{c}=F_{c}=F\alpha$ from the maximum permissible feedrate for the cornering angle and defined tolerance \eqref{eq: Tol 2 FIR, map}\eqref{eq: Tol 3 FIR, map}. (demonstrated in section \ref{Section:2 FIR Filters})
	\item Calculate pulse velocities $F$ and velocity blending pulse widths $T_{b}$ followed by modified velocity pulse widths $T_{v}$ \eqref{eq:Tv - single sideed}\eqref{eq:Tv main}.
	\item Synchronise timed axis velocity pulses and generate unfiltered axis velocity signals (pulse train).
	\item Define FIR filter time constants for the commanded feedrate from maximum permissible jerk \eqref{eq:Jerk Max}.
	\item Using high order FIR filtering with matching time constants interpolate the axis velocity pulse signals to generate smooth kinematic profiles for each axis \eqref{eq:Interpolate}.
	\item Finally, integrate the filtered velocity signals to generate synchronised accurate position commands in the time domain \eqref{eq: position integration}.
\end{enumerate}

Sections \ref{Section:Velocity Blending} and \ref{Section:Signal Generation} described the components of the velocity pulse train and application of FIR filtering for generation of kinematic profiles for non-stop high speed motion. The following sections will analytically demonstrate the relationship between the cornering speed $V_{c}$ to the blending error and axis kinematic limits and ultimately demonstrate how $F_{c}$ is selected to guarantee these constraints are satisfied. 

\subsection{Kinematic Profiles for the 2 First Order Filter Case} \label{Section:2 FIR Filters} 

The geometry of velocity blending pulses was presented and calculated in section \ref{Section:Velocity Blending}. The pulse signals are interpolated using FIR filters to generate kinematic profiles that control the cornering feedrate. This section analytically derives the equations for the kinematic profiles when using velocity blending pulses and FIR filtering based interpolation to control the cornering feedrate. In doing so, the authors are able to analytically calculate the blending pulse feedrate command $F_{c}$ which satisfies both TCP error and machine kinematic constraints during cornering transitions. 

\begin{figure}
		\centering
	\includegraphics[width=0.5\textwidth]{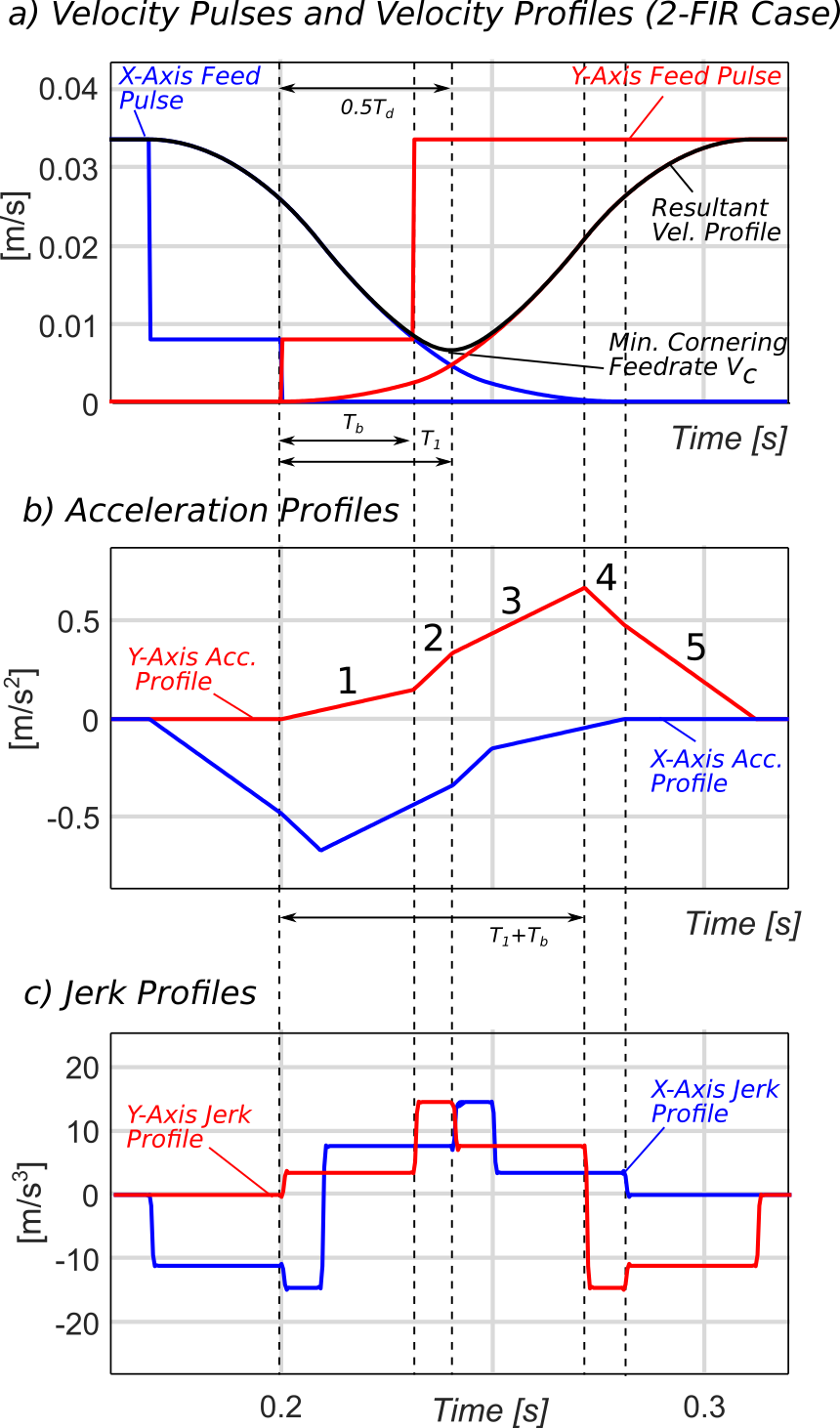}
	\caption{Interpolated kinematic profiles of velocity blending pulses using 2-FIR filters}
	\label{Fig:Kinematic Profiles Blending Pulse}
\end{figure}

Using 2-FIR filters with matching time constants to interpolate a velocity pulse signal results in the kinematic profiles shown in figure \ref{Fig:Kinematic Profiles Blending Pulse}. The profiles are split into 5 sections during acceleration/deceleration as shown in Fig.\ref{Fig:Kinematic Profiles Blending Pulse}b for the Y-axis acceleration. The objective of the analytical expressions is to calculate the interpolated displacement at the point of maximum TCP error and the interpolated velocity at the minimum cornering feedrate. This occurs at half the total filter delay $T_{d}/2$ (see Fig.\ref{Fig:Kinematic Profiles Blending Pulse}a). The total filter delay  for the 2-FIR case is $T_{d}=2T_{1}$, where $T_{1}$ is calculated from the maximum permissible jerk (equation \eqref{eq:Jerk Max}), therefore the maximum TCP error and minimum cornering feedrate occurs at $T_{1}$. Fig.\ref{Fig:Kinematic Profiles Blending Pulse}b shows $T_{1}$ is at the start of section 3, therefore only sections 1-3 of the kinematic profiles need considering. The analytical expressions for sections 1-3 of the displacement, velocity, acceleration and jerk profiles for the 2-FIR case are presented in equations \eqref{eq:FIR 2 TCP Disp Tb < T1} to \eqref{eq:FIR 2 TCP JerK Tb < T1} respectively. 

\begin{equation}\label{eq:FIR 2 TCP Disp Tb < T1}
s^{\prime}(t)=\left\{\begin{array}{ll}
\frac{1}{6}\frac{\alpha F}{T_{1}^2}t^3 & 0 \leq t<T_{b} \\
\frac{1}{6}\frac{F}{T_{1}^2}\left(t^3 +3T_{b}\left(\alpha-1\right)t^2 +3T_{b}\left(1-\alpha\right)t + T_{b}^3\left(\alpha-1\right)\right)  & T_{b} \leq t<T_{1}  \\
\frac{1}{6}\frac{F}{T_{1}^2}(\left(1-2\alpha\right)t^3 +\left(3T_{b}\left(\alpha-1\right)+6T_{1}\alpha\right)t^2 + (3T_{b}^2\left(1-\alpha\right)+\ldots \\ \ldots-6T_{1}^2 \alpha)t + T_{b}^3 \left(\alpha-1\right) + 2T_{1}^3 \alpha)  & T_{1} \leq t<T_{b}+T_{1} 
\end{array}\right.
\end{equation}

\begin{equation}\label{eq:FIR 2 TCP Vel Tb < T1}
v^{\prime}(t)=\left\{\begin{array}{ll}
\frac{1}{2}\frac{\alpha F}{T_{1}^2}t^2 & 0 \leq t<T_{b} \\
\frac{1}{2}\frac{F}{T_{1}^2}\left(t^2 +2T_{b}\left(\alpha-1\right)t-T_{b}^2\left(\alpha-1\right)\right) & T_{b} \leq t<T_{1}  \\
\frac{F}{2T_{1}^2}(\left(1-2\alpha\right)t^2 +\left(T_{1}\alpha + 2T_{b}\left(\alpha-1\right) +4T_{1}\alpha\right)t +T_{1}^2\alpha +\ldots\\\ldots+ T_{b}^2 \left(1-\alpha\right) - 2T_{1}\alpha) &
T_{1}\leq t<T_{b}+T_{1} 
\end{array}\right.\end{equation}

\begin{equation}\label{eq:FIR 2 TCP Acc Tb < T1}
a^{\prime}(t)=\left\{\begin{array}{ll}
\frac{\alpha F}{T_{1}^2}t & 0 \leq t<T_{b} \\
\frac{F}{T_{1}^2}\left(t+T_{b}\left(\alpha-1\right)\right) & T_{b} \leq t<T_{1}  \\
\frac{F}{T_{1}^2}\left(\left(1-2\alpha\right)t +2T_{1}\alpha -T_{b}+T_{b}\alpha\right) + & T_{1} \leq t<T_{b}+T_{1} 
\end{array}\right.\end{equation}

\begin{equation}\label{eq:FIR 2 TCP JerK Tb < T1}
j^{\prime}(t)=\left\{\begin{array}{ll}
\frac{\alpha F}{T_{1}^2} & 0 \leq t<T_{b} \\
\frac{F}{T_{1}^2} & T_{b} \leq t<T_{1}  \\
\frac{F}{T_{1}^2}\left(1-2\alpha\right) & T_{1} \leq t<T_{b}+T_{1} 
\end{array}\right.\end{equation}

The interpolated axis velocity at maximum TCP error (minimum cornering feedrate) occurs at $t = T_{d}/2 = T_{1}$, therefore in the 2-FIR filter case this results in the following expressions for interpolated velocity \eqref{eq: axis vel 2 TCP error} and displacement \eqref{eq: axis disp 2 TCP error}:
\begin{equation} \label{eq: axis vel 2 TCP error}
    v^\prime = \frac{1}{2}\frac{F}{T_{1}^2}\left(T_{1}^2 - T_{b}^2\left(1-\alpha\right)+2T_{1}T_{b}\left(\alpha-1\right)\right)
\end{equation}

\begin{equation} \label{eq: axis disp 2 TCP error}
    s^\prime = \frac{1}{6}\frac{F}{T_{1}^2}\left(T_{1}^3 + 3T_{1}T_{b}^2\left(1-\alpha\right)+3T_{1}^2T_{b}\left(\alpha-1\right) +T_{b}^3\left(\alpha-1\right)\right) 
\end{equation}

Using equation \eqref{Eq: Tb Equation}, the interpolated displacement \eqref{eq: axis vel 2 TCP error} and velocity\eqref{eq: axis disp 2 TCP error} can be expressed in terms of $F$ and $\alpha$:
\begin{equation}
v^\prime = \frac{F}{2}\,\alpha \,\left(-\alpha ^2+\alpha +1\right)
\end{equation}

\begin{equation} \label{eq: axis disp 2 TCP error alpha}
        s^\prime = \frac{F}{6}\,T_{1}\,\alpha \,\left(-\alpha ^3+\alpha ^2+1\right)
\end{equation}

\begin{figure}[ht]
		\centering
	\includegraphics[width=0.5\textwidth]{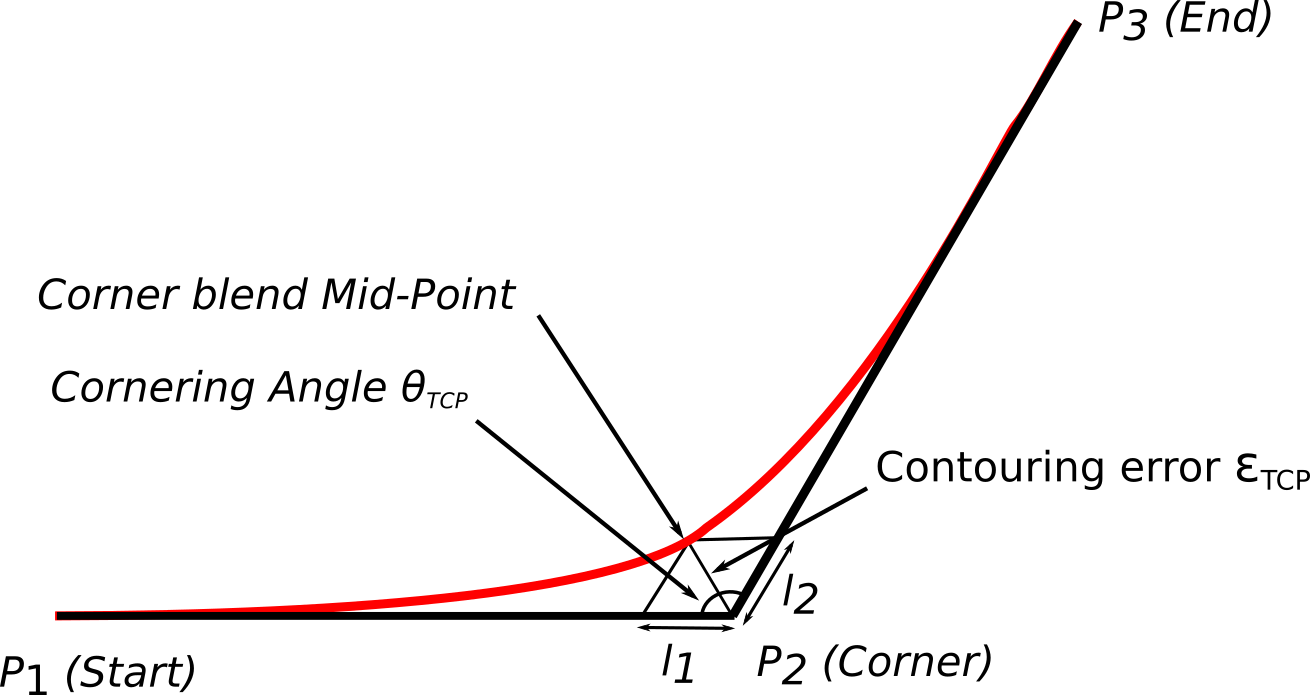}
	\caption{Toolpath showing contouring error and cornering angle between two consecutive G01 commands}
	\label{Fig:Cornering Angle}
\end{figure}

Fig.\ref{Fig:Cornering Angle} shows a cornering transition between two CL-lines or G01 commands. The maximum TCP contouring or corner blending error $\varepsilon_{T C P}$ occurs in the centre of the cornering trajectory and is calculated by evaluating the interpolated axis displacements $s^\prime$ at $t=T_{1}$. The interpolated axis displacements are calculated from \eqref{eq: axis disp 2 TCP error alpha} and the vectors from the corner transition to these positions are represented by $\mathbf{l_{1}}$ and $\mathbf{l_{2}}$.

The contouring error $\varepsilon_{T C P}$ (shown in Fig.\ref{Fig:Cornering Angle}) is calculated from the Euclidean distance between the vectors $\mathbf{l_{1}}$ and $\mathbf{l_{2}}$.

\begin{equation}\label{eq: FIR TCP s Tb < T1}
\varepsilon_{T C P}=\left\|\mathbf{l_{2}} -\mathbf{l_{1}} \overrightarrow{\mathrm{t}_{1}}\right\|=\sqrt{l_{1}^{2}+l_{2}^{2}+2 l_{1} l_{2}\cos{\theta_{TCP}}}
\end{equation}

where $\theta_{TCP}$ represents the TCP cornering angle. Assuming constant feedrate in this example, $l_{1} = l_{2} = l_{\varepsilon}$, in which case, \eqref{eq: FIR TCP s Tb < T1} simplifies to the following expression:

\begin{equation}\label{eq: FIR TCP calc 2}
\varepsilon_{T C P}^2\leq 2l_{\varepsilon}^{2}\left(1+\cos\theta_{TCP}\right)
\end{equation}

Inserting \eqref{Eq: Tb Equation} and \eqref{eq: axis disp 2 TCP error} into \eqref{eq: FIR TCP s Tb < T1} enables the TCP corner blending error to be defined as:

\begin{equation} \label{eq:Tol 2 FIR}
\varepsilon_{T C P}=\frac{\sqrt{2}}{6}\,\sqrt{F^2\,{T_{1}}^2\,\alpha ^2\,\left(\cos\theta_{TCP}+1\right)\,{\left(-\alpha ^3+\alpha ^2+1\right)}^2}
\end{equation}

Using equation \ref{eq:Tol 2 FIR} the TCP error can be calculated for any toolpath geometry and commanded feedrate. The kinematic profiles for the 3-FIR case are shown in Fig.\ref{fig:3_FIR_Kinematic_Profiles} in appendix \ref{Appendix 3-FIR Kinematic Profiles} and the derivation of TCP error for the 3-FIR case is included in appendix \ref{Appendix 3-FIR Filter Case}.

% Inspecting equation \ref{eq:Tol 2 FIR} the filter time constant can be calculated from the jerk parameter and commanded feedrate and so the only remaining variable is the feedrate scaling factor. For a given TCP error the feedrate scaling factor can be calculated. The next section will present how the equations for TCP error (for both 2-FIR and 3-FIR cases) can be used to calculate the maximum cornering feedrates that satisfy both tolerance and jerk constraints.

To ensure minimum cycle times the actual feedrate must remain as close to the commanded feedrate as possible throughout the toolpath including during cornering transitions. To satisfy both jerk and TCP error constraints, however, there is maximum permissible cornering feedrate. Using equations \eqref{eq:Tol 2 FIR} and \eqref{eq:Tol 3 FIR}, it is possible to calculate the relationship between TCP error, maximum permissible cornering feedrate and cornering angle for the 2-FIR and 3-FIR filter cases respectively.  

Rearranging equation \eqref{eq:Tol 2 FIR}, the maximum permissible cornering feedrate for the 2-FIR filter case must satisfy:
\begin{equation} \label{eq: Tol 2 FIR, map}
FT_{1}\sqrt{\cos\left(\theta_{TCP}+1\right)}\left(\alpha^4 -\alpha^3 -\alpha\right) -3\sqrt{2}\varepsilon_{TCP} \leq 0
\end{equation}

and for the 3-FIR filter using \eqref{eq:Tol 3 FIR} the maximum permissible cornering feedrate must satisfy:
\begin{equation}\label{eq: Tol 3 FIR, map}
FT_{1}\sqrt{\cos\left(\theta_{TCP}+1\right)}\left(16\alpha^5 + 16\alpha^4 -8\alpha^3 -16\alpha^2 -85\alpha - 1\right) - 192\sqrt{2}\varepsilon_{TCP}\leq 0
\end{equation}

For a commanded feedrate $F$ and range of cornering angles $\theta_{TCP}\in[0^{\circ},180^{\circ}]$, equations \eqref{eq: Tol 2 FIR, map} and \eqref{eq: Tol 3 FIR, map} are solved for solutions $0 \leq \alpha \leq 1 $ to calculate the limit to the feedrate scaling factor $\alpha$. When multiplied by the commanded feedrate $F$ this represents the maximum permissible cornering feedrate that can be achieved whilst satisfying the kinematic and tolerance constraints. The blending pulse feedrate $F_{c}$ is commanded to this limit value.  

\begin{figure}[ht]
		\centering
	\includegraphics[width=0.5\textwidth]{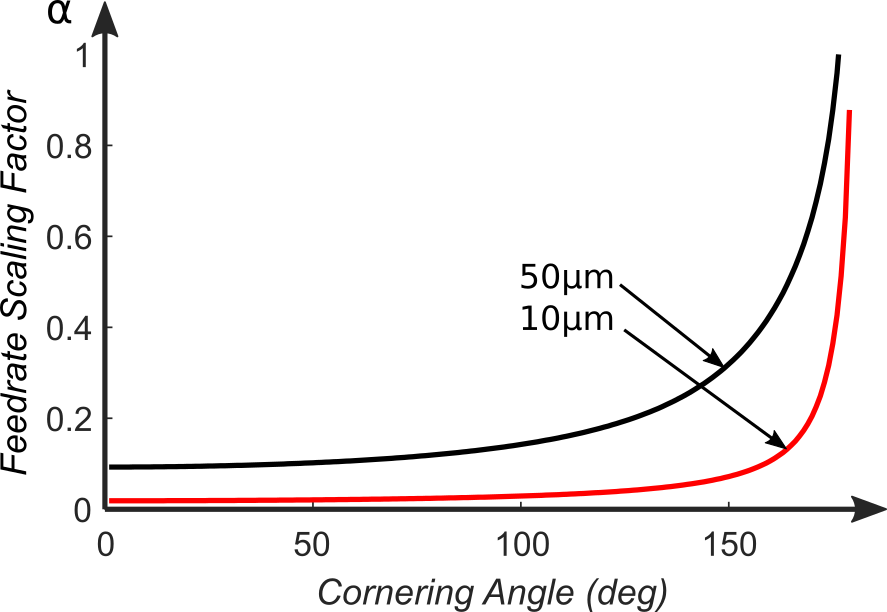}
	\caption{Minimum cornering feedrate and cornering angle curves shown for 10$\mu$m and 50$\mu$m tolerance settings at 2000 mm/min for 2-FIR case}
	\label{Fig:Corner Feedrate Relationship}
\end{figure}

The feedrate limit for the 2-FIR filter case is shown in Fig.\ref{Fig:Corner Feedrate Relationship}. Cornering feedrates selected below the curves will satisfy the TCP error constraints for the commanded feedrate and cornering angle. The figure shows the limits for 10$\mu$m and 50$\mu$m tolerance constraints. For the 50$\mu$m tolerance, the figure shows higher cornering feedrates can be achieved compared to the 10 $\mu$m case.  

\begin{figure}[ht]
		\centering
	\includegraphics[width=0.5\textwidth]{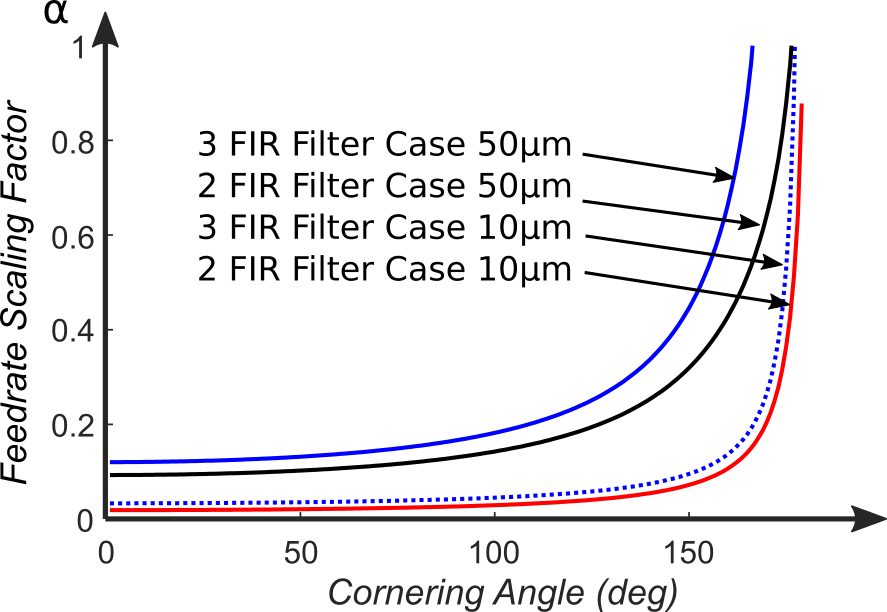}
	\caption{Minimum cornering feedrate and cornering angle curves shown for 10$\mu$m and 50$\mu$m tolerance settings at 2000 mm/min for both 2 and 3 first order FIR cases}
	\label{Fig:Corner Feedrate Comparison}
\end{figure}

The cornering feedrate limits for both the 2-FIR and 3-FIR filtered cases are compared in Fig.\ref{Fig:Corner Feedrate Comparison}. It can be seen that higher feedrates can be achieved in the 3-FIR case which satisfy the tolerance and jerk constraints. Therefore there is an advantage of using 3-FIR filters to reduce the overall machining cycle time as the tool can remain at higher feedrates during cornering transitions than for the 2-FIR case. Despite the advantage of using a higher order filter, there remains a limit to the order of filters that can be used effectively for trajectory generation. As the order is increased the filter time constant reduces. In the frequency domain the notch (as shown in Fig.\ref{Fig:Frequency Response}) will shift to higher frequencies. This will be constrained by the lowest structural mode of the machine tool.

This section has shown a method of using multiple first order FIR filters with matching time constants to model continuous linear interpolation of velocity pulse signals. It has been show that the cornering feedrate and TCP error can be controlled using velocity blending pulses. This method has been extended to predict feedrates and machining cycle time for toolpaths of any geometry and defined tolerance. The following section demonstrates and validates the proposed method on industrial case studies.   

\section{Experimental Validation}\label{Section:Validation}
Machining experiments were conducted on a DMG Mori Universal eVo 40 5-axis machining centre with a Heidenhain TNC640 controller. Two short tool-paths were used for pocketing operations and a single long aerospace part program is evaluated in the cycle time prediction. 

\subsection{Case Studies 1 \& 2 - Pocketing Toolpaths}
\begin{figure}[ht]
		\centering
	\includegraphics[width=0.5\textwidth]{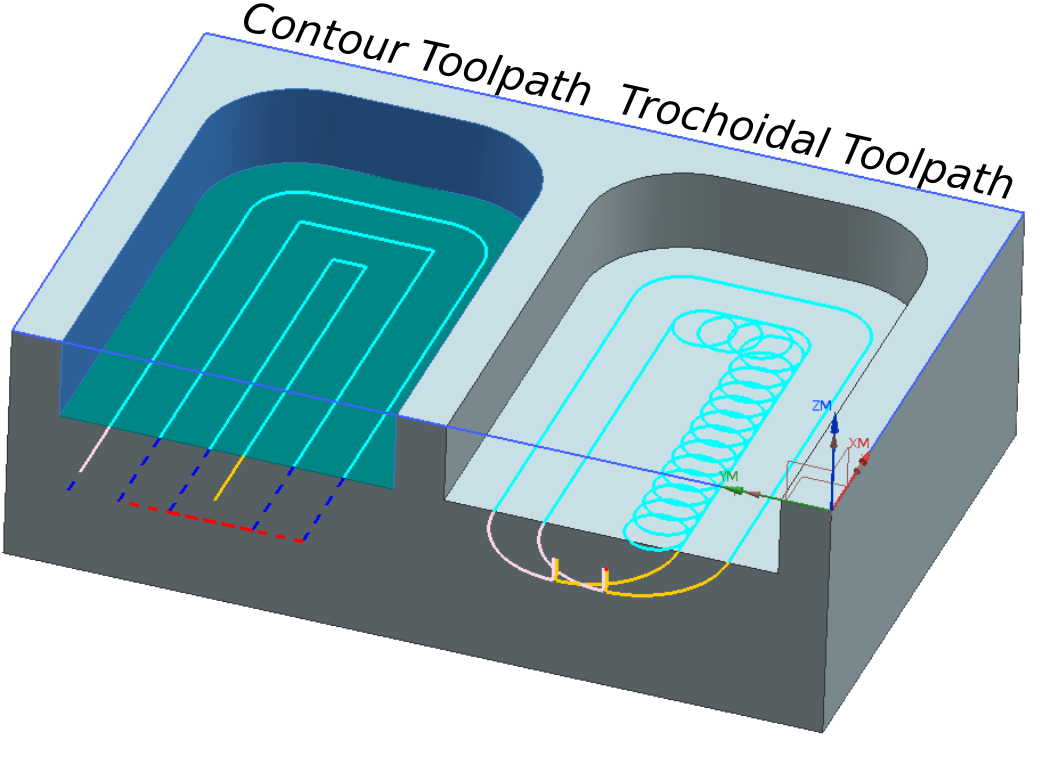}
	\caption{Contour (left) and Trochoidal (right) pocketing toolpaths designed in Siemens NX CAM}
	\label{Fig:NX Toolpaths}
\end{figure}

The first two case studies, as shown in Fig.\ref{Fig:NX Toolpaths}, consist of a contour and a trochoidal pocketing tool-path. These tool-paths are generated by CAM software \cite{SiemensPLMSoftware} and the part programs are deployed to the machine directly with no modification. Table \ref{Table:Cycle Time} shows the cutting conditions. As noted, 2 different feedrates 1000 and 3000 mm/min are used. The most important setting is contour error tolerance for HSM. Two different contouring tolerance, 10 and 50 $\mu$m are used. Table \ref{Table:Cycle Time} summarises the cycle time results. All simulated trajectories in the case studies were modelled using the method described in section \ref{Section:Signal Generation} and 3-FIR filters. 

\begin{figure}[ht]
		\centering
	\includegraphics[width=\textwidth]{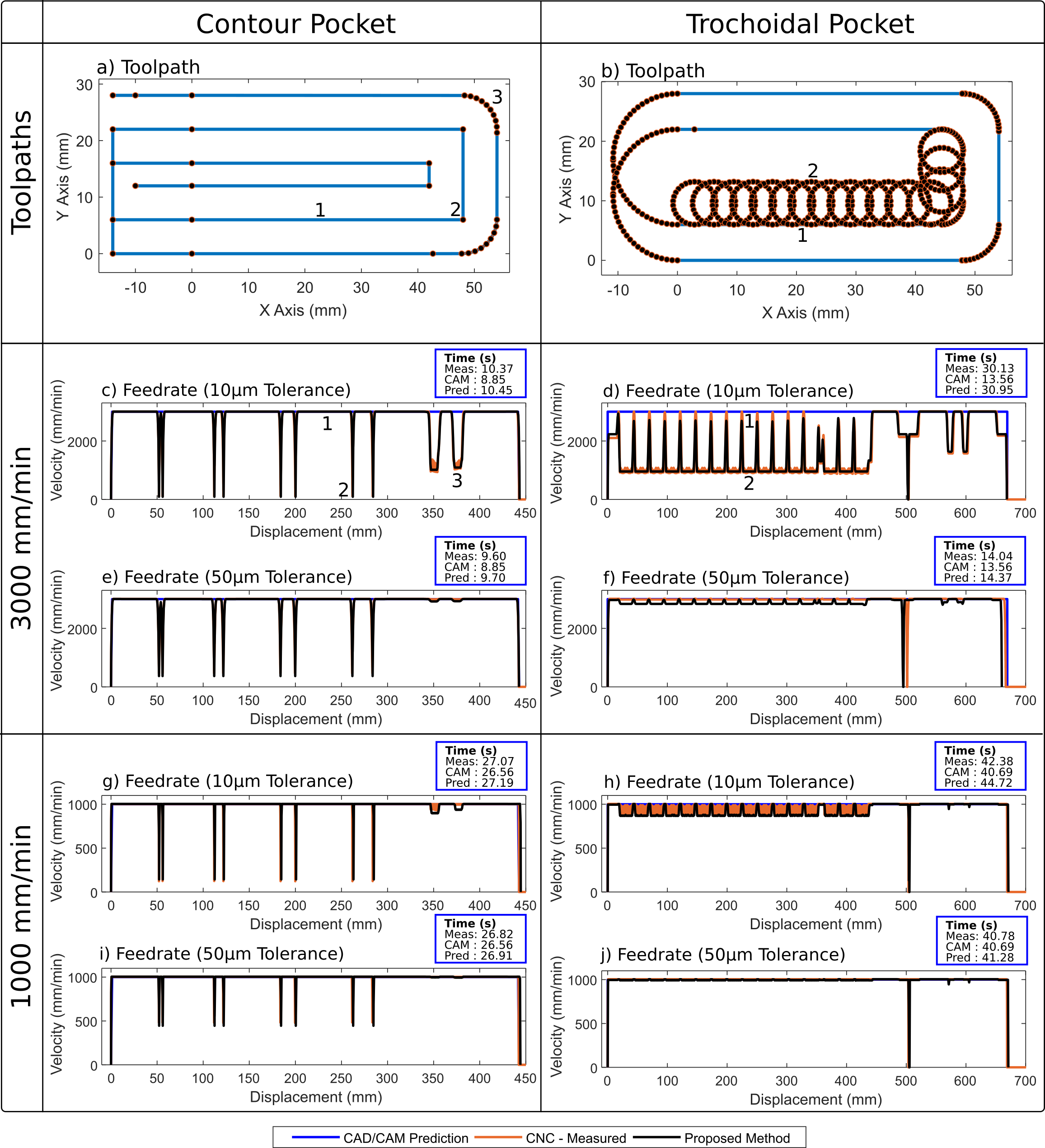}
	\caption{Pocketing case studies - predicted, measured and CAD/CAM tangential velocities}
	\label{Fig:Pocket Results}
\end{figure}

\subsubsection{Machining Cycle Time Estimation}

The predicted machining cycle times are compared with the measured CNC and CAD/CAM calculated machining cycle times. The results are presented in Table \ref{Table:Cycle Time} and Fig.\ref{Fig:Pocket Results}a. For all cases the predicted machining cycle times are accurate to within 3\% of the measured cycle time with the exception of the trochoidal pocket (1000 mm/min, 10$\mu$m case) which is 5.52\%. These compare favourably to the CAD/CAM calculated cycle times which has an error range from 0.22\% to 54.99\%.
The significant result is the Trochoidal pocket (3000 mm/min and 10$\mu$m case). The proposed method is able to accurately predict the increase in machining cycle time from 14.40 to 30.99 seconds when tightening the tolerance
from 50$\mu$m to 10$\mu$m, which is within 2.72\% of the measured cycle time. This is compared to an error of 54.99\% for the CAD/CAM calculated cycle time.
\begin{table}[ht]
	\begin{tabular}{|l|l|l|l|l|l|l|l|}
		\hline
		\textbf{Case Study} & \multicolumn{1}{c|}{\textbf{\begin{tabular}[c]{@{}c@{}}Feedrate\\ (mm/min)\end{tabular}}} & \multicolumn{1}{c|}{\textbf{\begin{tabular}[c]{@{}c@{}}Tolerance\\  (microns)\end{tabular}}} & \multicolumn{1}{c|}{\textbf{\begin{tabular}[c]{@{}c@{}}Measured \\ Time\\  (sec)\end{tabular}}} & \multicolumn{1}{c|}{\textbf{\begin{tabular}[c]{@{}c@{}}CAD/CAM \\ Time\\  (sec)\end{tabular}}} & \multicolumn{1}{c|}{\textbf{\begin{tabular}[c]{@{}c@{}}FIR\\Predicted\\  Time\\  (sec)\end{tabular}}} & \multicolumn{1}{c|}{\textbf{\begin{tabular}[c]{@{}c@{}}CAD/CAM\\  Prediction\\ Error (\%)\end{tabular}}} & \multicolumn{1}{c|}{\textbf{\begin{tabular}[c]{@{}c@{}}FIR \\  Prediction\\  Error (\%)\end{tabular}}} \\ \hline
		Contour Pocket      & 1000                                                                                      & 10                                                                                           & 27.07                                                                                           & 26.56                                                                                          & 27.32                                                                                            & -1.88                                                                                                     & 0.92                                                                                                         \\ \hline
		Contour Pocket      & 1000                                                                                      & 50                                                                                           & 26.82                                                                                           & 26.56                                                                                          & 26.95                                                                                            & -0.97                                                                                                     & 0.48                                                                                                         \\ \hline
		Contour Pocket      & 3000                                                                                      & 10                                                                                           & 10.37                                                                                           & 8.85                                                                                           & 10.46                                                                                            & -14.66                                                                                                    & 0.87                                                                                                         \\ \hline
		Contour Pocket      & 3000                                                                                      & 50                                                                                           & 9.6                                                                                             & 8.85                                                                                           & 9.73                                                                                              & -7.81                                                                                                     & 1.35                                                                                                         \\ \hline
		Trochoidal Pocket   & 1000                                                                                      & 10                                                                                           & 42.38                                                                                           & 40.69                                                                                          & 44.75                                                                                            & -3.98                                                                                                     & 5.59                                                                                                         \\ \hline
		Trochoidal Pocket   & 1000                                                                                      & 50                                                                                           & 40.78                                                                                           & 40.69                                                                                          & 41.30                                                                                            & -0.22                                                                                                     & 1.28                                                                                                         \\ \hline
		Trochoidal Pocket   & 3000                                                                                      & 10                                                                                           & 30.13                                                                                           & 13.56                                                                                          & 30.99                                                                                            & -54.99                                                                                                    & 2.86                                                                                                         \\ \hline
		Trochoidal Pocket   & 3000                                                                                      & 50                                                                                           & 14.04                                                                                           & 13.56                                                                                          & 14.40                                                                                            & -3.41                                                                                                     & 2.56                                                                                                         \\ \hline
	\end{tabular}
	\caption{Machining cycle times for contour and trochoidal pocket at different machining parameters}
	\label{Table:Cycle Time}
\end{table}

\subsubsection{Feedrate Prediction}

To demonstrate the performance of the feedrate prediction method a number of toolpath features were selected.
The predicted, CAD/CAM calculated and measured CNC tangential velocities at these particular features were recorded
and are presented in Tables 2 and 3. The contour pocket features consist of (1) a long G01 segment, (2) a sharp corner
and (3) a rounded corner consisting of small G01 segments. The trochoidal pocket features consist of (1) the stepover
segment and (2) the main arc. Depending on the tolerance and the commanded feedrate large differences in tangential
velocity can exist between the stepover segment and the main arc of a trochoidal toolpath which in turn results in a
large cyclical variation of cutting forces. It is for this reason they are included in this study.
The features described above are shown on the toolpaths in Fig. \ref{Fig:Pocket Results} and the corresponding position with respect
to displacement and tangential velocities are demonstrated directly beneath.
Overall, the prediction error ranges from 0.1-10.3\% compared with CAD/CAM calculated error range of 0.22-2555\%, where the error is calculated as a percentage difference from, and with respect to, the measured tangential velocity. The performance of the proposed feedrate prediction method at each feature is described below:

\paragraph{\textbf{Long G01 Segment.}} The prediction error range is between 0.1-0.13\% compared to the CAD/CAM calculated error range of 0.27-0.3\%. The high accuracy is to be expected as no feedrate limiting features are present in the segment. The difference in measured velocity compared to the idealised CAD/CAM values are due to interpolator rounding during trajectory generation.

\paragraph{\textbf{Sharp Corner.}}The prediction error range is between 7-20\% compared to the CAD/CAM calculated error range of 107-2555\%. The fundamental difference is due to the CAD/CAM calculation not taking into account the cornering kinematic constraints due to tolerance and thus not predicting the reduction in feedrate during the cornering segment. This holds true for all of the features demonstrated except the long G01 segment. For the 10$\mu$m tolerance cases the tool comes to an almost complete stop - 4\% and 11\% of the commanded feedrate for the 3000 mm/min and 1000 mm/min cases respectively, the presented method predicts these reductions.

\paragraph{\textbf{Rounded Corner.}} The prediction error range is between 0.3-5.4\% compared to the CAD/CAM calculated error range of 0.2-168\%. The significant result is the 3000 mm/min and 10$\mu$m case (Fig. \ref{Fig:Pocket Results}c) where the CAD/CAM calculation
does not account for the reduction in velocity due to the tolerance requirement. The CAD/CAM calculated error is 168\% compared to the measured value and the prediction error is within 2.4\%.

\paragraph{\textbf{Trochoid Stepover.}} The prediction error range is between 0.1-10\% compared to the CAD/CAM prediction error range of 0.2-0.5\%. The CAD/CAM calculation does not predict any differences along the trochoidal toolpath between the stepover and the main arc. This can be seen in Fig. \ref{Fig:Pocket Results}d for the 3000 mm/min 10$\mu$m case. The blue line shows the CAD/CAM prediction but the actual kinematic profile is very different. The stepover results in tangential velocities close to the commanded feedrate as the cornering angles between the segments are less acute than for the rest of the main arc.

\paragraph{\textbf{Trochoid Main Arc.}} The prediction error range is between 0.6-7\% compared to the CAD/CAM calculated error range of 0.2-208\%. The reduction in tangential velocity around the main arc is due to the cornering angles between the segments. The influence of the toolpath tolerance on the cornering tangential velocity can be seen in Fig.\ref{Fig:Pocket Results}d
and Fig.\ref{Fig:Pocket Results}f. The increase in tolerance from 50$\mu$m to 10$\mu$m results in more than a 65\% reduction in tangential velocity
around the main arcs of the trochoids. The prediction method accurately predicts the feedrate within 1.5\% of tangential velocity measured at the main arc. Taking this result one step further, this demonstrates that a feedrate driven cutting force model when incorporating the prediction method will be able to predict the cyclical cutting forces due to the 65\% variation in magnitude of feedrate fluctuations around the trochoidal toolpath.

% Please add the following required packages to your document preamble:
% \usepackage{multirow}
\begin{table}[ht]
	\begin{tabular}{|l|c|c|c|c|l|c|c|c|}
		\hline
		\multicolumn{1}{|c|}{\multirow{2}{*}{\textbf{\begin{tabular}[c]{@{}c@{}}Case \\ Study\end{tabular}}}} & \multirow{2}{*}{\textbf{\begin{tabular}[c]{@{}c@{}}Feedrate\\  (mm/min)\end{tabular}}} & \multirow{2}{*}{\textbf{\begin{tabular}[c]{@{}c@{}}Tolerance\\  (microns)\end{tabular}}} & \multirow{2}{*}{\textbf{\begin{tabular}[c]{@{}c@{}}Analysis\\  Point\end{tabular}}}   & \multicolumn{5}{c|}{\textbf{Tangential Velocities (mm/min)}}                                                                                                                                                              \\ \cline{5-9} 
		\multicolumn{1}{|c|}{}                                                                                &                                                                                        &                                                                                          &                                                                                       & \textbf{Measured} & \multicolumn{1}{c|}{\textbf{CAD/CAM}} & \textbf{\begin{tabular}[c]{@{}c@{}}Proposed\\ Prediction\end{tabular}} & \textbf{\begin{tabular}[c]{@{}c@{}}CAD/CAM\\ Error\end{tabular}} & \textbf{\begin{tabular}[c]{@{}c@{}}Proposed\\Prediction\\  Error\end{tabular}} \\ \hline
		Contour                                                                                               & 1000                                                                                   & 10                                                                                       & \multirow{4}{*}{\begin{tabular}[c]{@{}c@{}}Point 1\\ Straight\\ G01\end{tabular}}     & 1003              & 1000                                  & 1004                & -0.30\%                                                          & +0.10\%                                                              \\ \cline{1-3} \cline{5-9} 
		Contour                                                                                               & 1000                                                                                   & 50                                                                                       &                                                                                       & 1003              & 1000                                  & 1004                & -0.30\%                                                          & +0.10\%                                                              \\ \cline{1-3} \cline{5-9} 
		Contour                                                                                               & 3000                                                                                   & 10                                                                                       &                                                                                       & 3008              & 3000                                  & 3005                & -0.27\%                                                          & -0.10\%                                                              \\ \cline{1-3} \cline{5-9} 
		Contour                                                                                               & 3000                                                                                   & 50                                                                                       &                                                                                       & 3008              & 3000                                  & 3004                & -0.27\%                                                          & -0.13\%                                                              \\ \hline
		Contour                                                                                               & 1000                                                                                   & 10                                                                                       & \multirow{4}{*}{\begin{tabular}[c]{@{}c@{}}Point 2\\ Sharp\\ Corner\end{tabular}}     & 124               & 1000                                  & 143                 & +706\%                                                           & +15\%                                                                \\ \cline{1-3} \cline{5-9} 
		Contour                                                                                               & 1000                                                                                   & 50                                                                                       &                                                                                       & 482               & 1000                                  & 446                 & +107\%                                                           & -7\%                                                                 \\ \cline{1-3} \cline{5-9} 
		Contour                                                                                               & 3000                                                                                   & 10                                                                                       &                                                                                       & 113               & 3000                                  & 93                  & +2555\%                                                          & -17\%                                                                \\ \cline{1-3} \cline{5-9} 
		Contour                                                                                               & 3000                                                                                   & 50                                                                                       &                                                                                       & 465               & 3000                                  & 370                 & +545\%                                                           & -20\%                                                                \\ \hline
		Contour                                                                                               & 1000                                                                                   & 10                                                                                       & \multirow{4}{*}{\begin{tabular}[c]{@{}c@{}}Point 3\\  Rounded\\  Corner\end{tabular}} & 994               & 1000                                  & 940                 & +0.6\%                                                           & -5.4\%                                                               \\ \cline{1-3} \cline{5-9} 
		Contour                                                                                               & 1000                                                                                   & 50                                                                                       &                                                                                       & 998               & 1000                                  & 995                 & +0.2\%                                                           & -0.3\%                                                               \\ \cline{1-3} \cline{5-9} 
		Contour                                                                                               & 3000                                                                                   & 10                                                                                       &                                                                                       & 1120              & 3000                                  & 1093                & +168\%                                                           & -2.4\%                                                               \\ \cline{1-3} \cline{5-9} 
		Contour                                                                                               & 3000                                                                                   & 50                                                                                       &                                                                                       & 2996              & 3000                                  & 2929                & +0.13\%                                                          & -2.23\%                                                              \\ \hline
	\end{tabular}
	\caption{Contour pocket case study: tangential velocity prediction and performance}
	\label{Table:contour velocitye}
\end{table}

\begin{table}[ht]
	\begin{tabular}{|l|c|c|c|c|c|c|c|c|}
		\hline
		\multicolumn{1}{|c|}{\multirow{2}{*}{\textbf{\begin{tabular}[c]{@{}c@{}}Case Study   \\ Pocket\end{tabular}}}} & \multirow{2}{*}{\textbf{\begin{tabular}[c]{@{}c@{}}Feedrate   \\  (mm/min)\end{tabular}}} & \multirow{2}{*}{\textbf{\begin{tabular}[c]{@{}c@{}}Tolerance\\  (microns)\end{tabular}}} & \multirow{2}{*}{\textbf{\begin{tabular}[c]{@{}c@{}}Analysis\\  Point\end{tabular}}}    & \multicolumn{5}{c|}{\textbf{Tangential Velocities (mm/min)}}                                                                                                                                           \\ \cline{5-9} 
		\multicolumn{1}{|c|}{}                                                                                         &                                                                                           &                                                                                          &                                                                                        & \textbf{Measured} & \textbf{CAD/CAM} &  \textbf{\begin{tabular}[c]{@{}c@{}}Proposed \\ Prediction\end{tabular}} & \textbf{\begin{tabular}[c]{@{}c@{}}CAD/CAM \\  Error\end{tabular}} & \textbf{\begin{tabular}[c]{@{}c@{}}Proposed\\Prediction\\  Error\end{tabular}} \\ \hline
		Trochoidal                                                                                                     & 1000                                                                                      & 10                                                                                       & \multirow{4}{*}{\begin{tabular}[c]{@{}c@{}}Point 1\\ Trochoid\\ Stepover\end{tabular}} & 1005              & 1000             & 1003                & -0.50\%                                                            & 0.20\%                                                               \\ \cline{1-3} \cline{5-9} 
		Trochoidal                                                                                                     & 1000                                                                                      & 50                                                                                       &                                                                                        & 1005              & 1000             & 1004                & -0.50\%                                                            & 0.10\%                                                               \\ \cline{1-3} \cline{5-9} 
		Trochoidal                                                                                                     & 3000                                                                                      & 10                                                                                       &                                                                                        & 2994              & 3000             & 2687                & +0.20\%                                                            & 10.3\%                                                               \\ \cline{1-3} \cline{5-9} 
		Trochoidal                                                                                                     & 3000                                                                                      & 50                                                                                       &                                                                                        & 3014              & 3000             & 2985                & -0.46\%                                                            & 0.96\%                                                               \\ \hline
		Trochoidal                                                                                                     & 1000                                                                                      & 10                                                                                       & \multirow{4}{*}{\begin{tabular}[c]{@{}c@{}}Point 2\\ Trochoid\\ Main Arc\end{tabular}} & 938               & 1000             & 872                 & +6.61\%                                                            & 7\%                                                                  \\ \cline{1-3} \cline{5-9} 
		Trochoidal                                                                                                     & 1000                                                                                      & 50                                                                                       &                                                                                        & 1002              & 1000             & 994                 & -0.20\%                                                            & 0.60\%                                                               \\ \cline{1-3} \cline{5-9} 
		Trochoidal                                                                                                     & 3000                                                                                      & 10                                                                                       &                                                                                        & 973               & 3000             & 958                 & +208\%                                                             & 1.5\%                                                                \\ \cline{1-3} \cline{5-9} 
		Trochoidal                                                                                                     & 3000                                                                                      & 50                                                                                       &                                                                                        & 2986              & 3000             & 2838                & +0.47\%                                                            & 4.96\%                                                               \\ \hline
	\end{tabular}
	\caption{Trochoidal pocket case study: tangential velocity prediction and performance }
	\label{Table:trochoidal velocit}
\end{table}

\subsection{Case Study 3 - Aerostructure Toolpath}
\begin{figure}
		\centering[ht]
	\includegraphics[width=0.7\textwidth]{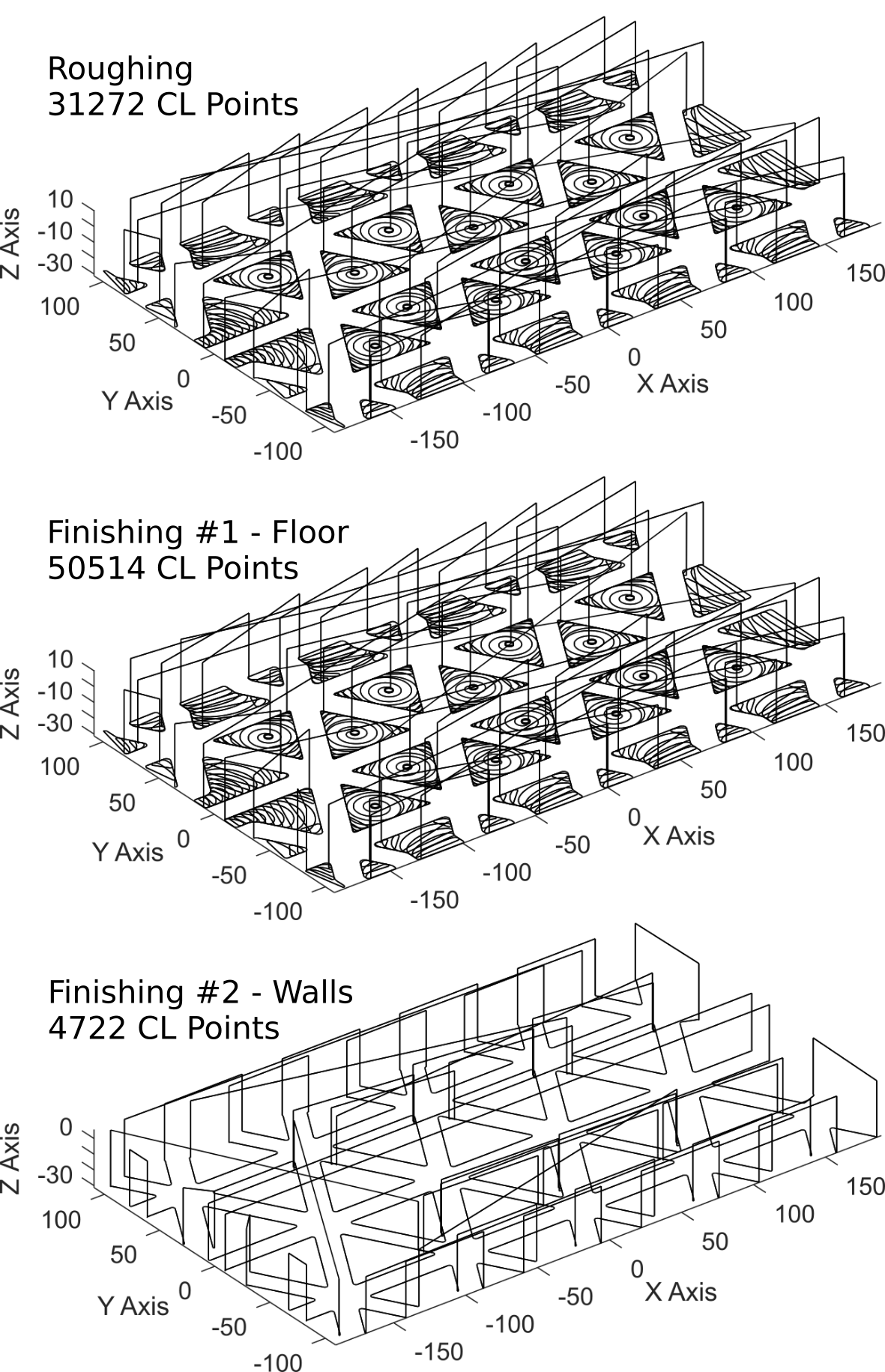}
	\caption{Aerostructure toolpaths (shown in order of operation)}
	\label{Fig:Aerostructure Toolpath}
\end{figure}

An industrial toolpath was chosen to validate the method against a representative aerostructure part. The part program consists of three toolpaths - roughing, finishing \#1 floors and finishing \#2 walls as shown in Fig. \ref{Fig:Aerostructure Toolpath}. The part programs were run at three tolerance settings, 10$\mu$m, 20$\mu$m and 50$\mu$m to demonstrate the significant impact tolerance has on machining cycle times and therefore on feedrate and cycle time prediction. Table \ref{table:cycle time aero} compares the predicted machining cycle times with both the measured cycle times and the predicted times from a commercial CAD/CAM software package for each individual toolpath. 

% \begin{figure}
% 		\centering
% 	\includegraphics[width=0.5\textwidth]{ISO_aerostructure_part.jpg}
% 	\caption{Representative Aerostructure Component Post-Machining}
% 	\label{Fig:Aerostructure Photo}
% \end{figure}

The overall machining cycle times, calculated by summing the cycle times for the 3 sections of the part program, are shown in Table \ref{Table: Total Times}. The CAD/CAM prediction error ranges from 62.41\% under prediction for the 10$\mu$m case to 36.42\% under prediction for the 50$\mu$m case. The actual CAD/CAM predicted times do not change as the software does not account for tolerance, the calculation is based upon distance travelled along the toolpath and ideal feedrate. Therefore as the tolerance is relaxed the measured cycle time approaches the CAD/CAM case and their prediction becomes more accurate. 

The prediction error from the proposed method (as shown in Table \ref{Table: Total Times}) ranges from 3.50\% over prediction for the 10$\mu$m case to 4.69\% for the 50$\mu$m case. The 20$\mu$m case has a prediction error of 5.34\% under the measured cycle time which is approximately 10\% of the CAD/CAM error (51.49\%) for that particular case. The aerostructure case study validates the model for predicting both feedrate and machining cycle times for varying tolerance settings on very complex industrial toolpaths far outperforming the CAM software.

\begin{table}[ht]
	\begin{tabular}{|c|c|c|c|c|c|c|c|}
		\hline
		\textbf{Section} & \textbf{\begin{tabular}[c]{@{}c@{}}Feedrate\\  (mm/min)\end{tabular}} & \textbf{\begin{tabular}[c]{@{}c@{}}Tolerance\\  (microns)\end{tabular}} & \textbf{\begin{tabular}[c]{@{}c@{}}Measured \\ Time\\ (sec)\end{tabular}} & \textbf{\begin{tabular}[c]{@{}c@{}}CAD/CAM\\ Time\\ (sec)\end{tabular}} & \textbf{\begin{tabular}[c]{@{}c@{}}Proposed\\Prediction \\ Time\\ (sec)\end{tabular}} & \textbf{\begin{tabular}[c]{@{}c@{}}CAD/CAM\\ Error (\%)\end{tabular}} & \textbf{\begin{tabular}[c]{@{}c@{}}Proposed\\ Prediction\\    Error (\%)\end{tabular}} \\ \hline
		Roughing         & 8000                                                                  & 10                                                                      & 1017.40                                                                   & 332                                                                     & 1032.80                                                                    & -67.37                                                                                & 1.51                                                                                     \\ \hline
		Finish Floor     & 8000                                                                  & 10                                                                      & 543.75                                                                    & 213                                                                     & 606.75                                                                     & -60.83                                                                                & 11.59                                                                                    \\ \hline
		Finish Walls     & 8000                                                                  & 10                                                                      & 133.31                                                                    & 92                                                                      & 114.30                                                                     & -30.99                                                                                & -14.26                                                                                   \\ \hline
		Roughing         & 8000                                                                  & 20                                                                      & 752.40                                                                    & 332                                                                     & 711.05                                                                     & -55.87                                                                                & -5.50                                                                                    \\ \hline
		Finish Floor     & 8000                                                                  & 20                                                                      & 435.01                                                                    & 213                                                                     & 420.85                                                                     & -51.04                                                                                & -3.26                                                                                    \\ \hline
		Finish Walls     & 8000                                                                  & 20                                                                      & 125.61                                                                    & 92                                                                      & 111.01                                                                     & -26.76                                                                                & -11.62                                                                                   \\ \hline
		Roughing         & 8000                                                                  & 50                                                                      & 551.52                                                                    & 332                                                                     & 478.66                                                                     & -39.80                                                                                & -13.21                                                                                   \\ \hline
		Finish Floor     & 8000                                                                  & 50                                                                      & 334.37                                                                    & 213                                                                     & 368.28                                                                     & -36.30                                                                                & 10.14                                                                                    \\ \hline
		Finish Walls     & 8000                                                                  & 50                                                                      & 115.92                                                                    & 92                                                                      & 107.90                                                                     & -20.63                                                                                & -6.92                                                                                    \\ \hline
	\end{tabular}
	\caption{Machining cycle time comparison for Aerostructure part case study.}
	\label{table:cycle time aero}
\end{table}

\begin{table}[ht]
	\begin{tabular}{ccc|c|c|c|c|c|}
		\cline{4-8}
		\multicolumn{1}{l}{}                & \multicolumn{1}{l}{}                                                                      & \multicolumn{1}{l|}{}                                                  & \multicolumn{3}{r|}{\textbf{Total Machining Cycle Times}}                                                                                                                                                 & \multicolumn{2}{c|}{\textbf{Errors (\%)}}                                                                                                             \\ \hline
		\multicolumn{1}{|c|}{\textbf{Case}} & \multicolumn{1}{c|}{\textbf{\begin{tabular}[c]{@{}c@{}}Feedrate\\ (mm/min)\end{tabular}}} & \textbf{\begin{tabular}[c]{@{}c@{}}Tolerance\\ (microns)\end{tabular}} & \textbf{\begin{tabular}[c]{@{}c@{}}Measured\\ (sec)\end{tabular}} & \textbf{\begin{tabular}[c]{@{}c@{}}CAD/CAM\\ (sec)\end{tabular}} & \textbf{\begin{tabular}[c]{@{}c@{}}Proposed \\Prediction\\ (sec)\end{tabular}} & \textbf{\begin{tabular}[c]{@{}c@{}}CAD/CAM\\ \end{tabular}} & \textbf{\begin{tabular}[c]{@{}c@{}}Proposed\\ Prediction\end{tabular}} \\ \hline
		\multicolumn{1}{|c|}{1}             & \multicolumn{1}{c|}{8000}                                                                 & 10                                                                     & 1694.46                                                           & 637                                                              & 1753.85                                                            & -62.41                                                                   & 3.50                                                                       \\ \hline
		\multicolumn{1}{|c|}{2}             & \multicolumn{1}{c|}{8000}                                                                 & 20                                                                     & 1313.02                                                           & 637                                                              & 1242.91                                                            & -51.49                                                                   & -5.34                                                                      \\ \hline
		\multicolumn{1}{|c|}{3}             & \multicolumn{1}{c|}{8000}                                                                 & 50                                                                     & 1001.81                                                           & 637                                                              & 954.84                                                             & -36.42                                                                   & -4.69                                                                      \\ \hline
	\end{tabular}
	\caption{Total machining cycle times and errors for measured, predicted and CAD/CAM.}
	\label{Table: Total Times}
\end{table}

\newpage

\section{Cutting Force Prediction} \label{Section:Cutting Force}

Lastly, the importance of accurate feedrate prediction for virtual machining models is demonstrated. This is realized by estimating cutting forces along the complex trochoidal toolpath shown in Fig.\ref{Fig:Cutting Forces}. Predicting the cutting forces, considering the complex tool engagements on this toolpath, is realized by adapting the cutting force prediction model presented in \cite{Berglind2017} with the proposed feedrate prediction method. Readers should refer to \cite{Armendia2019} and \cite{Berglind2017} for details of the cutting force model.

\newpage
\subsection{Case Study 4 - Accurate Cutting Force Prediction using Predicted Feedrates}\label{Section: Cutting Force Prediction}

\begin{figure}
	\centering[ht]
	\includegraphics[width=0.4\textwidth]{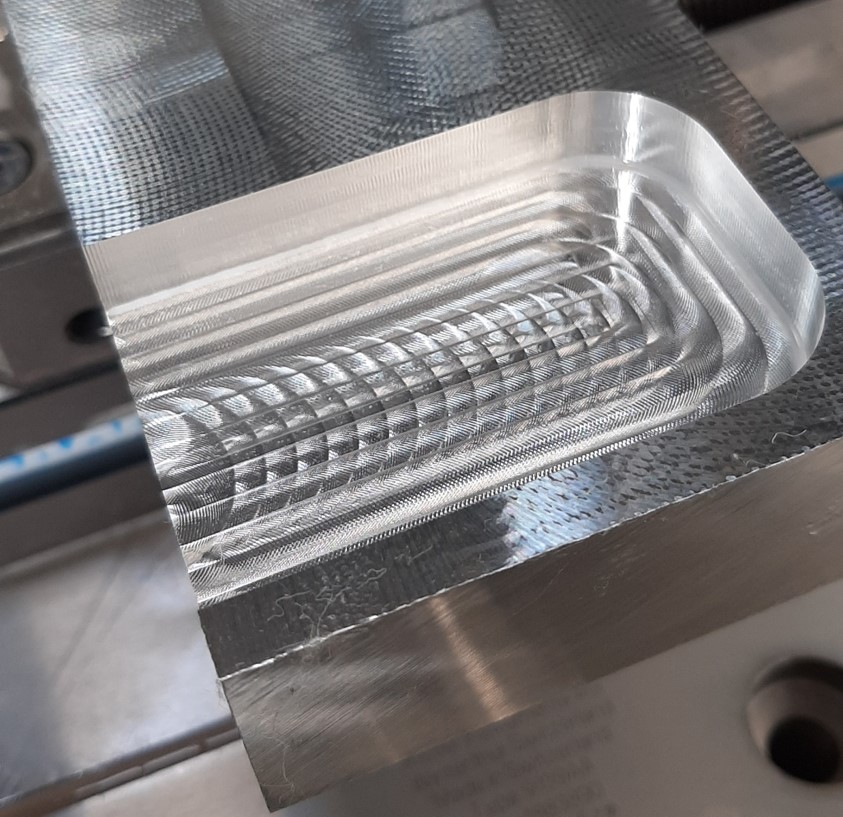}
	\caption{Machined AL7075 pocket using a trochoidal toolpath}
	\label{Fig:Roughed Pocket}
\end{figure}

\begin{figure}[ht]
	\centering
	\includegraphics[width=\textwidth]{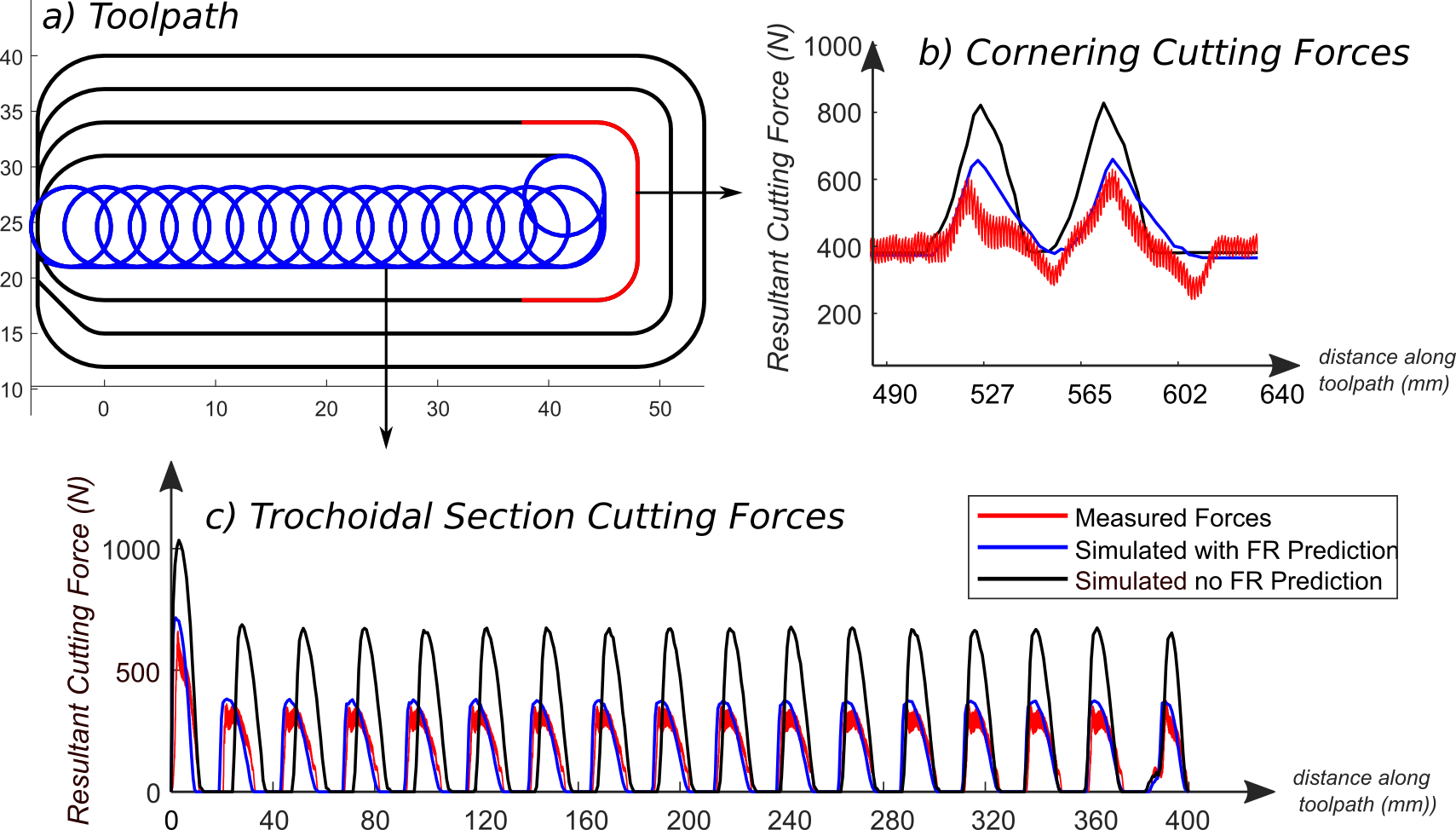}
	\caption{Simulated and measured resultant cutting forces for a trochoidal pocket}
	\label{Fig:Cutting Forces}
\end{figure}

To validate the feedrate prediction method with a cutting force model machining trials were conducted on the 5-axis DMG Mori eVo 40 machining centre fitted with a Heidenhain TNC640 controller.
The toolpath, shown in Fig. \ref{Fig:Cutting Forces}, was designed using NX CAM as a trochoidal pocketing operation. A 40mm x 60mm x 10mm open sided pocket was selected as the test feature as shown in Fig. \ref{Fig:Roughed Pocket}. A 2-fluted 12mm solid carbide end mill with a HSK-63A tool holder was used. The workpieces were 236mm x 30mm x 6mm aluminium 7075, each held using a Geradi compact grip vice mounted to the dynamometer. A Kistler 9139AA dynamometer and a National Instruments USB-6343 multi-channel DAQ was used to acquire cutting force data at 10kHz. The machining centre was connected to a local area network via a RJ45 network cable such that the machine controller data was accessed by two methods. The first using a pre-defined MTConnect datastream through a TCPIP connection at 20Hz and the second using an LSV2 protocol direct to the controller through a TCPIP connection at 111Hz.

The predicted cutting forces during the trochoidal section are shown in Fig. \ref{Fig:Cutting Forces}. The peak predicted cutting force for the standard feedrate model is 673N compared to 380N for the filtered feedrate model, from the peak measured cutting forces this gives prediction errors of 96.2\% and 10.8\% respectively. In the cornering section of the toolpath the peak predicted cutting force for the standard feedrate model is 821N compared to 656N  for the filtered feedrate model, from the peak measured cutting forces this gives prediction errors of 37.3\% and 9.7\% respectively. The validation trials show that the inclusion of an accurate feedrate profile in the cutting force model enables a more accurate prediction of cutting forces for complex toolpaths.

\section{Conclusions} \label{Section:Conclusion}

A novel method of accurately modelling the trajectory generation of NC systems has been proposed. The main conclusions from this research are as follows: 

\begin{enumerate}
	\item An accurate method of feedrate prediction along short-segmented complex tool-paths was introduced.
	\item The linear interpolation dynamics and commanded axis kinematic profiles of NC systems were predicted using both 2 and 3 first order Finite Impulse Response filters with the same time constant.
	\item The corner blending behaviour during non-stop interpolation of linear segments was modeled by introducing velocity blending pulses.
	\item For the first time, the minimum cornering feedrate, that satisfies both the tolerance and machining constraints, has been calculated analytically for toolpaths of any geometry.
	\item The reduction in machining cycle time by using 3 FIR filters compared to 2 FIR filters was proven analytically.
	\item  The feedrate prediction method was validated experimentally against four different case studies demonstrating industrial 3-axis machining tool-paths. 
	\item The proposed method demonstrated cycle times can be estimated with $>$90\% accuracy, greatly outperforming CAM-based predictions.  
	\item The predicted feedrate method was incorporated into a cutting force model, demonstrating an increase in cutting force accuracy for a complex toolpath, and validated experimentally. 
\end{enumerate}

Further work will integrate the methods into virtual machining and digital-twin models and extend the method to 5-axis machining.

\appendix
\newpage

\section{2-FIR Filters with Matching Time Constants}\label{Appendix Filter Matching}
\begin{figure}[ht]
        \centering
    \includegraphics[width=0.35\textwidth]{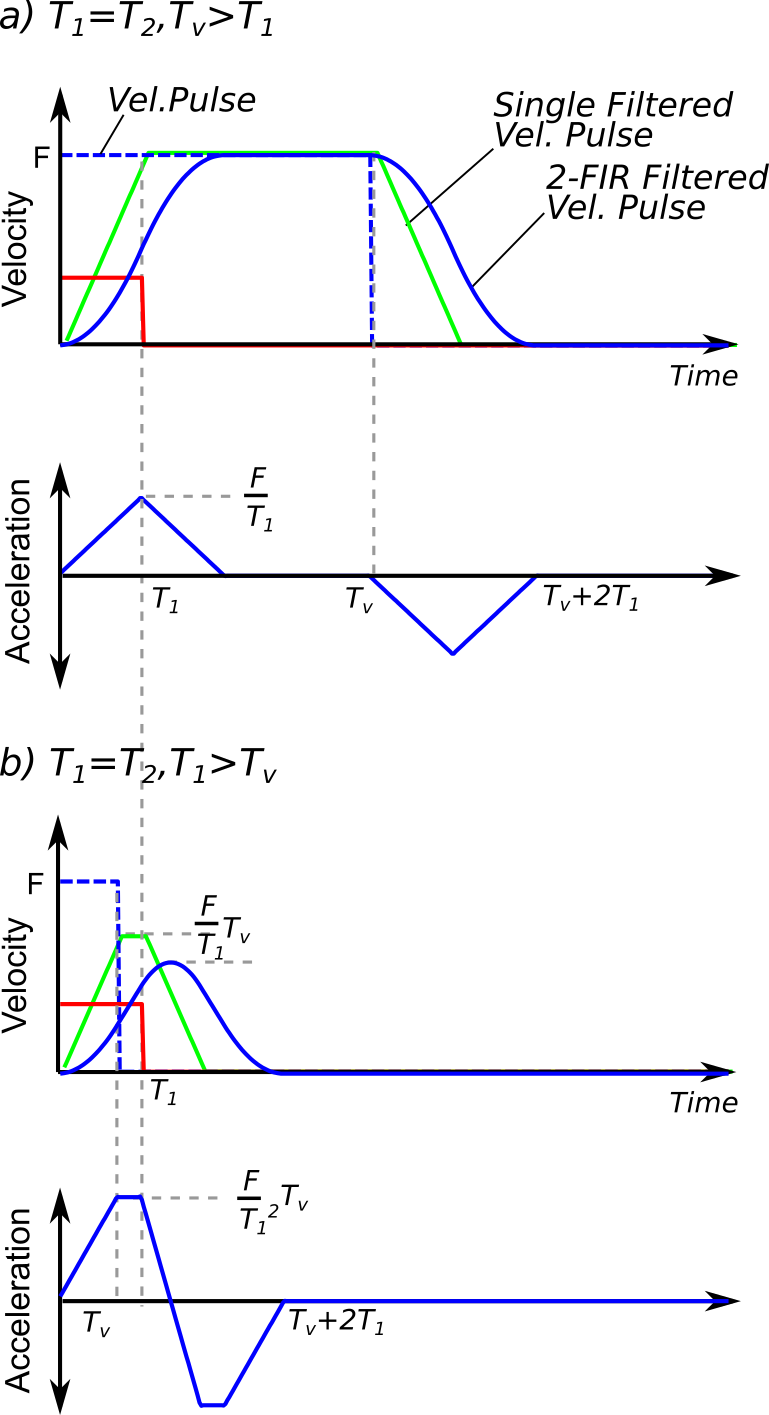}
    \caption{Velocity and acceleration profiles generated by 2 FIR filters with matching time constants.}
    \label{fig:Filters_matching}
\end{figure}

\newpage
\section{Kinematic Equations for 3-FIR Case}\label{Appendix 3 Filter Equations}

% 3 Filter Equations

\begin{equation}\label{eq:FIR 3 T1 Vel}
v^{\prime}(t)=\left\{\begin{array}{ll}
\frac{F}{6T_{1}^{3}}t^3 & 0 \leq t<T_{1} \\
\frac{F}{2T_{1}^3}\left(-\frac{2}{3}t^3+3T_{1}t^2-3T_{1}^2t+T_{1}^3\right) & T_{1} \leq t<2T_{1} \\
\frac{F}{2T_{1}^3}\left(\frac{1}{3}t^3-3T_{1}t^2+9T_{1}^2t-7T_{1}^3
\right) & 2T_{1} \leq t<3T_{1} \\
F & 3T_{1} \leq t<T_{v} \\
\frac{F}{2T_{1}^3}\left(-\frac{1}{3}t^3+T_{v}t^2 -T_{v}t +\frac{1}{3}T_{v}^3 +2T_{1}^3\right) & T_{v} \leq t<T_{v} + T_{1} \\
\frac{F}{2T_{1}^3}\left(\frac{2}{3}t^3-2T_{v}t^2 -3T_{1}t^2+3T_{1}^2t+2T_{v}^2t +6T_{v}T_{1}t -\frac{2}{3}T_{v}^3-3T_{v}^2T_{1}-3T_{v}T_{1}^2 +T_{1}^3\right) & T_{v}+T_{1} \leq t<T_{v}+2T_{1} \\
\frac{F}{2T_{1}^3}\left(-\frac{1}{3}t^3 +T_{v}t^2 +3T_{1}t^2 -T_{v}^2 t -9T_{1}^2 t -6T_{v}T_{1}t +\frac{1}{3}\left(T_{v}+3T_{1}\right)^3\right) & T_{v}+2T_{1} \leq t<T_{v}+3T_{1} 
\end{array}\right.\end{equation}

\begin{equation}\label{eq:FIR 3 T1 Acc}
a^{\prime}(t)=\left\{\begin{array}{ll}
\frac{F}{2T_{1}^{3}}t^2 & 0 \leq t<T_{1} \\
\frac{F}{T_{1}^3}\left(-t^2+3T_{1}t-\frac{3}{2}T_{1}^2\right) & T_{1} \leq t<2T_{1} \\
\frac{F}{T_{1}^3}\left(\frac{1}{2}t^2-3T_{1}t+\frac{9}{2}T_{1}^2\right) & 2T_{1} \leq t<3T_{1} \\
0 & 3T_{1} \leq t<T_{v} \\
\frac{F}{T_{1}^3}\left(-\frac{1}{2}t^2+T_{v}t-\frac{1}{2}T_{v}^2\right) & T_{v} \leq t<T_{v} + T_{1} \\
\frac{F}{T_{1}^3}\left(t^2-2T_{v}t -3T_{1}t+\
\frac{3}{2}T_{1}^2+T_{v}^2+3T_{v}T_{1}\right) & T_{v}+T_{1} \leq t<T_{v}+2T_{1} \\
\frac{F}{2T_{1}^3}\left(-t^2+2T_{v}t+6T_{1}t-T_{v}^2-9T_{1}^2 -6T_{v}T_{1}\right) & T_{v}+2T_{1} \leq t<T_{v}+3T_{1} 
\end{array}\right.\end{equation}

\begin{equation}\label{eq:FIR 3 T1 Jerk}
j^{\prime}(t)=\left\{\begin{array}{ll}
\frac{F}{T_{1}^{3}}t & 0 \leq t<T_{1} \\
\frac{2F}{T_{1}^3}\left(-t + \frac{3}{2}T_{1}\right) & T_{1} \leq t<2T_{1} \\
\frac{F}{T_{1}^3}\left(t -3T_{1}\right) & 2T_{1} \leq t<3T_{1} \\
0 & 3T_{1} \leq t<T_{v} \\
\frac{F}{T_{1}^3}\left(-t+T_{v}\right) & T_{v} \leq t<T_{v} + T_{1} \\
\frac{2F}{T_{1}^3}\left(t-T_{v}-\frac{3}{2}T_{1}\right) & T_{v}+T_{1} \leq t<T_{v}+2T_{1} \\
\frac{F}{T_{1}^3}\left(-t+T_{v}+3T_{1}\right) & T_{v}+2T_{1} \leq t<T_{v}+3T_{1} 
\end{array}\right.\end{equation}

\newpage
\section{Kinematic Profiles for 3-FIR Case}\label{Appendix 3-FIR Kinematic Profiles}

\begin{figure}[ht]
    \centering
    \includegraphics[width = 0.5\textwidth]{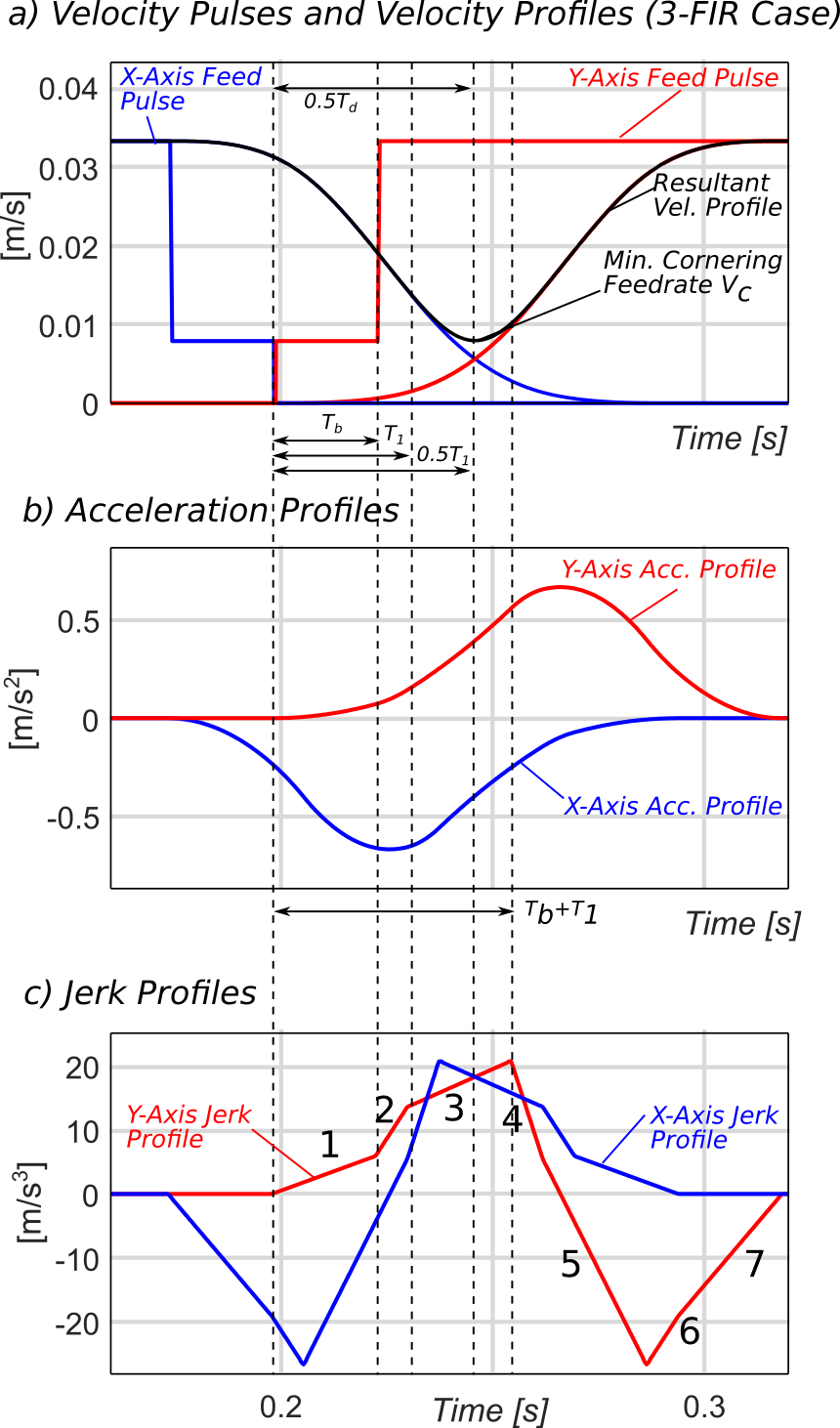}
    \caption{Velocity, acceleration and jerk profiles generated by blended velocity pulses interpolated 3-FIR filters}
    \label{fig:3_FIR_Kinematic_Profiles}
\end{figure}

\newpage

\section{Kinematic Equations for the 3-FIR Filter Case (Blending Pulses)}\label{Appendix 3-FIR Filter Case}

% As shown in section \ref{Section:2 FIR Filters} the TCP error can be analytically derived from interpolating velocity blending pulses with FIR filters. This section introduces an additional first order FIR filter and derives the TCP error equation. 

% When using 3 first order FIR filters to interpolate a velocity blending pulse the total filter delay is $T_{d}=3T_{1}$. The filtered kinematic profiles is split into 6 segments as seen in Fig. \ref{Fig:Kinematic Profiles Blending Pulse}. Equations \eqref{eq:FIR 3 TCP Displacement Tb < T1},\eqref{eq:FIR 3 TCP Vel Tb < T1}, \eqref{eq:FIR 3 TCP Acc Tb < T1} and \eqref{eq:FIR 3 TCP JerK Tb < T1} represent the analytical expressions for the first 3 kinematic sections of the displacement, velocity, acceleration and jerk respectively. 

\begin{equation}\label{eq:FIR 3 TCP Displacement Tb < T1}
s^{\prime}(t)=\left\{\begin{array}{ll}
\frac{\alpha F}{24T_{1}^3}t^4 & 0 \leq t<T_{b} \\
\frac{2\alpha F}{T_{1}^3}\left(-\frac{1}{24}t^4+\frac{1}{4}T_{1}t^3-\frac{3}{8}T_{1}^2t^2\right)+\frac{\alpha F}{2}t -\frac{\alpha FT_{1}}{8}^2 & T_{b} \leq t<T_{`} \\
\frac{F\ \left(1-3\alpha\right)}{24T_{1}^3}\left(-t^4+ 4T_{b}t^3 -6T_{b}^2 t^2 +4T_{b}^3 t -T_{b}^4\right) + \frac{\alpha F}{24T_{1}^3}(-8b_{1}t^3 +18T_{1}^2 +12T_{b}^2 + \ldots \\ \ldots -36T_{1}T_{b} +24T_{b})t^2 + \left(8T_{b}^2 -12T_{1}^3\right)t+3T_{1}^4-18T_{b}^4 +24T_{1}T_{b}^3 -16Tb^3b_{1} & T_{1} \leq t<T_{1}+T_{b}
\end{array}\right.
\end{equation}

\begin{equation}\label{eq:FIR 3 TCP Vel Tb < T1}
v^{\prime}(t)=\left\{\begin{array}{ll}
\frac{\alpha F}{6T_{1}^3}t^3 & 0 \leq t<T_{b} \\
\frac{2\alpha F}{T_{1}^3}\left(-\frac{1}{6}t^3+\frac{3}{4}T_{1}t^2-\frac{3}{4}T_{1}^2t\right)+\frac{\alpha F}{2} & T_{b} \leq t<T_{1}  \\
  \frac{F\left(1-3\alpha\right)}{6T_{1}^3}\left(t^3 -3T_{b}t^2 +3T_{b}^2t +T_{b}^3\right) +\frac{\alpha F}{T_{1}^3}\left(\frac{1}{2}T_{1}^3-\frac{1}{3}T_{b}^3-(\frac{3}{2}T_{1}^2-3T_{1}T_{b}+T_{b}^2\right)t+\ldots \\ \ldots+b_{1}t^2 - 2T_{b}b_{1}t) & T_{1} \leq t<T_{1}+T_{b} 
\end{array}\right.\end{equation}

\begin{equation}\label{eq:FIR 3 TCP Acc Tb < T1}
a^{\prime}(t)=\left\{\begin{array}{ll}
\frac{\alpha F}{T_{1}^3}t & 0 \leq t<T_{b} \\
\frac{2\alpha F}{T_{1}^3}\left(-\frac{1}{2}t^2+\frac{3}{2}T_{1}t -\frac{3}{4}T_{1}^2\right) & T_{b} \leq t<T_{1}  \\
\frac{2\alpha F}{T_{1}^3}\left(\frac{3}{2}T_{1}^2-\frac{3}{2}T_{1}T_{b}+\frac{1}{2}T_{b}^2-T_{b}b_{1}+b_{1}\right) - \frac{F\left(1-3\alpha\right)}{T_{1}^3}\left(\frac{1}{2}T_{b}^2 - \left(T_{b}t-\frac{1}{2}t^2\right)\right) & T_{1} \leq t<T_{1}+T_{b} 
\end{array}\right.\end{equation}

\begin{equation}\label{eq:FIR 3 TCP JerK Tb < T1}
j^{\prime}(t)=\left\{\begin{array}{ll}
\frac{\alpha F}{T_{1}^3}t & 0 \leq t<T_{b} \\
\frac{\alpha F}{T_{1}^3}\left(-t+\frac{3}{2}T_{1}\right) & T_{b} \leq t<T_{1}  \\
\frac{F\left(1-3\alpha\right)}{T_{1}^3}\left(t-T_{b}\right) + \frac{2\alpha F}{T_{1}^3}\left(\frac{3}{2}T_{1}-T_{b}\right) & T_{1} \leq t<T_{1}+T_{b} 
\end{array}\right.\end{equation}

where $b_{1} = \frac{3}{2}T_{1}-T_{b}$.

The maximum TCP error occurs at $t = \frac{T_{d}}{2} = \frac{3}{2}T_{1}$ for the 3 first order FIR filter case, the interpolated axis velocity and displacement, are defined as equations \eqref{eq: axis disp 3 TCP error} and \eqref{eq:FIR 3 TCP Disp at 1.5T1 Tb > T1} respectively:

\begin{equation} \label{eq: axis disp 3 TCP error}
v^\prime  =\frac{F}{48\,{T_{1}}^3}\,\left(36\,T_{1}T_{b}^2-54\,T_{1}^2\,T_{b}-3\,T_{1}^3\,\alpha +8\,T_{b}^3\,\alpha +27\,T_{1}^3-8\,T_{b}^3-36\,T_{1}\,T_{b}^2\,\alpha +54\,T_{1}^2\,T_{b}\,\alpha \right)
\end{equation}

\begin{equation}\label{eq:FIR 3 TCP Disp at 1.5T1 Tb > T1}
\begin{split}
s^\prime=-\frac{F}{384T_{1}^3}\,(96\,T_{1}\,T_{b}^3+216\,T_{1}^3\,T_{b}+3\,T_{1}^4\,\alpha +16\,T_{b}^4\,\alpha -81\,T_{1}^4- -16\,T_{b}^4-216\,T_{1}^2\,T_{b}^2+...\\ ...+216\,T_{1}^2\,T_{b}^2\,\alpha -96\,T_{1}\,T_{b}^3\,\alpha -216\,T_{1}^3\,T_{b}\,\alpha)
\end{split}
\end{equation}

Using equation \eqref{Eq: Tb Equation}, \eqref{eq: axis disp 3 TCP error} and \eqref{eq:FIR 3 TCP Disp at 1.5T1 Tb > T1} can be expressed in terms of $F$ and $\alpha$ as:
\begin{equation}
    v^\prime = \frac{F}{48}\left(- 8\alpha^4 - 4\alpha^3 + 6\alpha^2 + 29\alpha + 1\right)
\end{equation}

\begin{equation}
    s^\prime = \frac{F}{384}T_{1}\left(- 16\alpha^5 - 16\alpha^4 + 8\alpha^3 + 16\alpha^2 + 85\alpha + 1\right)
\end{equation}

Solving equation \eqref{eq: FIR TCP calc 2} with \eqref{eq:FIR 3 TCP Disp at 1.5T1 Tb > T1} results in the maximum TCP error for the 3-FIR filter case as follows:

\begin{equation}\label{eq:Tol 3 FIR}
\varepsilon_{T C P}=\frac{\sqrt{2}}{384}\,\sqrt{F^2\,{T_{1}}^2\,\left(\cos\theta_{TCP}+1\right)\,{\left(-16\,\alpha ^5-16\,\alpha ^4+8\,\alpha ^3+16\,\alpha ^2+85\,\alpha +1\right)}^2}
\end{equation}

\begin{acknowledgements}
This research was supported by the Advanced Manufacturing Research Centre members and EPSRC (grant EP/L016257/1).
\end{acknowledgements}

\section{Conflict of interest/ Competing Interests}
The authors have no conflicts of interest to declare that are relevant to the content of this article.
\section{Consent to Participate}
Not applicable
\section{Consent to Publish}
The authors declare that they all consent to publication.
\section{Ethical Approval}
Not applicable
\section{Funding}
This research was supported by the Advanced Manufacturing Research Centre members and EPSRC (grant EP/L016257/1).

\section{Availability of data and materials}
Data are available Rob Ward with the permission of University of Sheffield AMRC. The data that support the findings of this study are available from the corresponding author,RW, upon reasonable request.

\bibliographystyle{IEEEtranN}

\end{document}